\documentclass[a4paper,fleqn]{cas-dc}

\usepackage[numbers]{natbib}
\usepackage{subfig}
\usepackage{amsmath}
\usepackage{lineno}  % use \linenumbers
\usepackage{siunitx} % degree symbol 
\usepackage{multicol}
\usepackage{multirow}
\usepackage{soul}
\begin{document}
\let\WriteBookmarks\relax
\def\floatpagepagefraction{1}
\def\textpagefraction{.001}
\shorttitle{pedestrian contact interaction trajectory characterization}
\shortauthors{J. Kwak~\textit{et~al}.}

\title[mode = title]{Characterizing pedestrian contact interaction trajectories to understand spreading risk in human crowds}         
%\tnotemark[1,2]
% Analysing pedestrian contact interaction to understand spreading risk in human crowds.
% Understanding the relation between pedestrian contact interaction and epidemic spreading: A statistical testing approach.
%
% Authors
%
\author[1]{Jaeyoung Kwak}
\cormark[1]
%\cortext[cor1]{Corresponding author: Jaeyoung Kwak (jaeyoung.kwak@ntu.edu.sg)}
%%\ead{jaeyoung.kwak@ntu.edu.sg}
\author[2]{Michael H. Lees}
\author[1]{Wentong Cai}

\address[1]{School of Computer Science and Engineering, Nanyang Technological University, 50 Nanyang Avenue, Singapore 639798, Singapore}
\address[2]{Informatics Institute, University of Amsterdam, Science Park 904, Amsterdam 1098XH, The Netherlands}

\begin{abstract}
A spreading process can be observed when particular information, substances, or diseases spread through a population over time in social and biological systems. It is widely believed that contact interactions among individual entities play an essential role in the spreading process. Although contact interactions are often influenced by geometrical conditions, little attention has been paid to understand their effects, especially on contact duration among pedestrians. To examine how the pedestrian flow setups affect contact duration distribution, we have analyzed trajectories of pedestrians in contact interactions collected from pedestrian flow experiments of uni-, bi- and multi-directional setups. Based on turning angle entropy and efficiency, we have classified the type of motion observed in the contact interactions. We have found that the majority of contact interactions in the unidirectional flow setup can be categorized as confined motion, hinting at the possibility of long-lived contact duration. However, ballistic motion is more frequently observed in the other flow conditions, yielding frequent, brief contact interactions. Our results demonstrate that observing more confined motions is likely associated with the increase of parallel contact interactions regardless of pedestrian flow setups. This study highlights that the confined motions tend to yield longer contact duration, suggesting that the infectious disease transmission risk would be considerable even for low transmissibility. These results have important implications for crowd management in the context of minimizing spreading risk.

This work is an extended version of Kwak~\textit{et~al}.~(2023) presented at the 2023 International Conference on Computational Science (ICCS).

\end{abstract}

%\includegraphics{figs/grabs.pdf}
%\item Research highlights item 1
%\item Research highlights item 2
%\item Research highlights item 3
\maketitle

%\linenumbers
\section{Introduction}
\label{section:intro}
Modeling contact interactions among individual entities is essential to understand spreading processes in social and biological systems, such as information diffusion in human populations~\cite{Samar_IEEE2006,Wu_IEEE2008} and transmission of infectious disease in animal and human groups~\cite{Hu_2013,Manlove_2022}. For the spreading processes in social and biological systems, one can observe a contact interaction when two individual entities are within a close distance, so they can exchange substance and information or transmit disease from one to the other one. In previous studies, macroscopic patterns of contact interactions are often estimated based on simple random walking behaviors, including ballistic motion. For example, Rast~\cite{Rast_PRE2022} simulated continuous-space-time random walks based on ballistic motion of non-interacting random walkers. Although such random walk models have widely applied to estimate contact duration for human contact networks, little work has been done to study the influence of pedestrian flow geometrical conditions on the distribution of contact duration.

To examine how the geometrical conditions of pedestrian flow affect the contact duration distribution, we perform trajectory analysis for the experimental dataset collected from a series of experiments performed for various pedestrian flow setups. The trajectory analysis of moving organisms, including proteins in living cells, animals in nature, and humans, has been a popular research topic in various fields such as biophysics~\cite{Saxton_1997,Manzo_2015,Shen_2017}, movement ecology~\cite{Benhamou_2007,Edelhoff_2016,Getz_PNAS2008}, and epidemiology~\cite{Rutten_SciRep2022,Wilber_2022}. Single particle tracking (SPT) analysis, a popular trajectory analysis approach frequently applied in biophysics and its neighboring disciplines, characterizes the movement dynamics of individual entities based on observed trajectories~\cite{Manzo_2015,Qian_1991}. According to SPT analysis, one can identify different types of diffusion, for instance, directed diffusion in which individuals move in a clear path and confined diffusion in which individuals tend to move around the initial position. The most common method for identifying diffusion types is based on the mean-squared displacement (MSD), which reflects the deviation of an individual's position with respect to the initial position after time lag~\cite{Qian_1991,Michalet_PRE2010}. Motion types can be identified based on the diffusion exponent. MSD has been widely applied for various trajectory analysis studies in biophysics~\cite{Goulian_2000,Hubicka_PRE2020}. For pedestrian flow trajectory analysis, Murakami~\textit{et~al.}~\cite{Murakami_interface2019,Murakami_SciAdv2021} analyzed experimental data of bidirectional pedestrian flow and reported diffusive motion in individual movements perpendicular to the flow direction. They suggested that uncertainty in predicting neighbors' future motion contributes to the appearance of diffusive motion in pedestrian flow.

Previous studies have demonstrated the usefulness of SPT analysis in examining the movement of individuals. However, SPT analysis does not explicitly consider relative motion among individuals in contact, suggesting that analyzing the relative motion can reveal patterns that might not be noticeable from the SPT analysis approach. For example, if two nearby individuals are walking in parallel with a similar speed, one might be able to see various shapes of relative motion trajectories, although the individual trajectories are nearly straight lines. For contact interaction analysis, the analysis of relative motion trajectories can be utilized to predict the length of contact duration and identify contact interaction characteristics, such as when the interacting individuals change walking direction significantly. Regarding the spreading processes, examining relative motion trajectories can be used to identify optimal geometrical conditions that can minimize contact duration when diseases are being transmitted through face-to-face (direct) interactions. 

Although MSD is simple to apply, MSD has limitations. Refs.~\cite{Briane_2016,Briane_PRE2018,Janczura_PRE2020} pointed out that MSD might not be suitable for short trajectories to extract meaningful information. Additionally, due to its power-law form, the estimation of the diffusion exponent in MSD is prone to estimation errors~\cite{Kepten_PLOS2015,Burnecki_SciRep2015}. As an alternative to MSD, various approaches have been proposed, including the statistical test approach~\cite{Briane_2016,Briane_PRE2018,Weron_PRE2019,Janczura_2022} and machine learning approaches~\cite{Kowalek_JPhysA2022}.

In our resent work~\cite{Kwak_iccs2023}, we analyzed pedestrian contact interaction trajectories based on a statistical testing approach. For the statistical test procedure, we measured a standardized value of the largest distance traveled by individuals from their starting point during the contact interaction. Ref.~\cite{Kwak_iccs2023} demonstrated potential in identifying different types of contact interaction observed from experimental data including uni-, bi-, and multi-directional flow~\cite{Holl_Dissertation2016,Cao_JSTAT2017,url_dataset}.

In this work, we present an improved version of the pedestrian contact interaction trajectory analysis studied in Ref.~\cite{Kwak_iccs2023}. To capture a wide range of contact interactions, we have generated synthetic trajectories based on random walk models whose characteristics are well understood. While Ref.~\cite{Kwak_iccs2023} uses one test statistic for contact interaction trajectory characterization, this study applies a set of test statistics in an attempt to systematically characterize contact interaction trajectories for different pedestrian flow conditions. According to the previous studies~\cite{Briane_2016,Briane_PRE2018,Janczura_PRE2020}, we have quantified the classification criteria for the contact interaction trajectories.

The remainder of this paper is organized as follows. Section~\ref{section:methods} presents our approach for contact interaction trajectory classification in line with statistical testing approach. In that section, we describe the test procedure including synthetic trajectory generation, evaluation, and classification criteria identification. In Section~\ref{section:results}, we briefly summarize the dataset used for our analysis and then discuss the findings of our analysis in terms of contact interaction trajectory classification and contact durations.

\section{Methods}
\label{section:methods}
\begin{figure*}
	\centering
	\includegraphics[width=14cm]{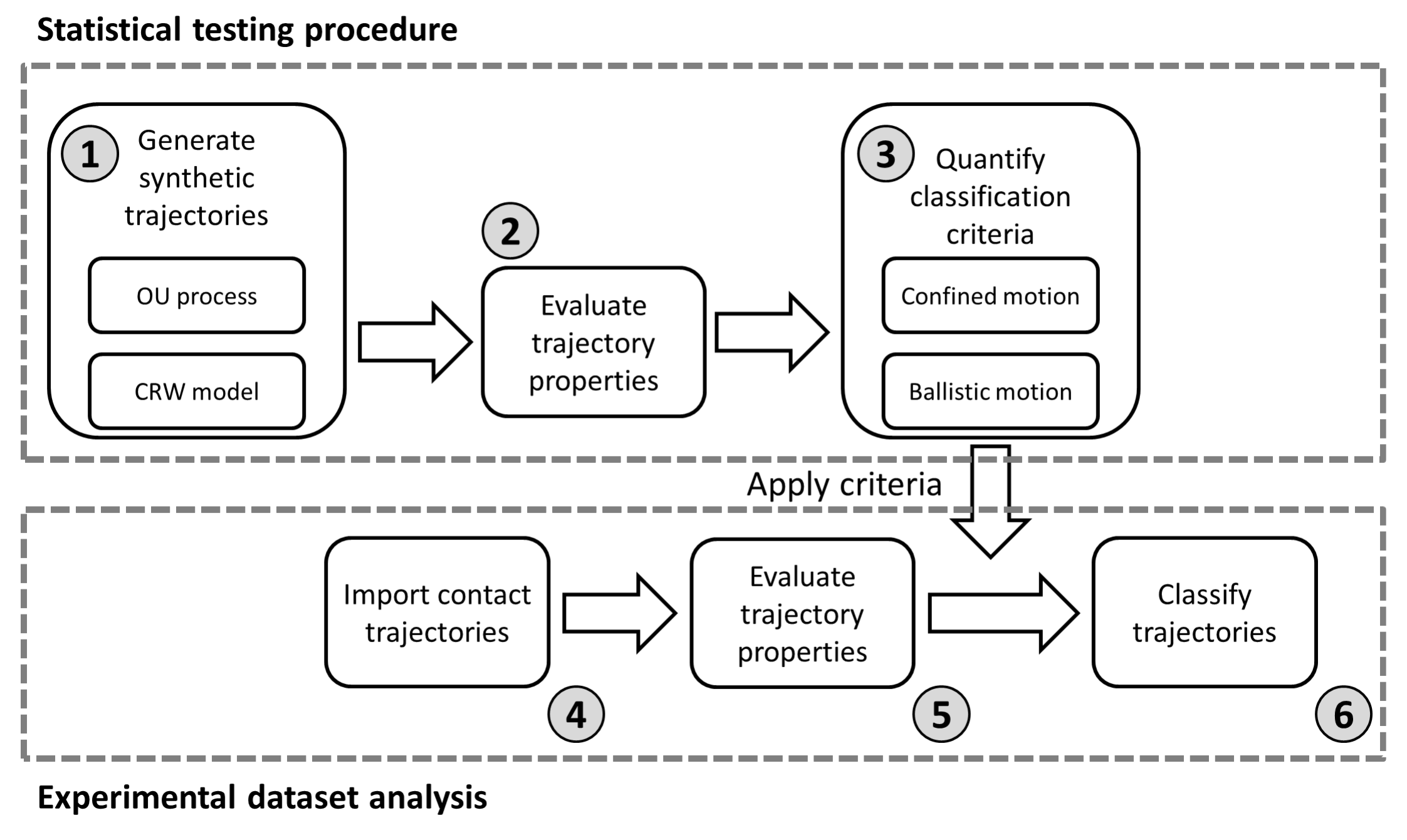}
	\caption{Schematic representation of our approach for contact interaction trajectory classification: (step 1) generate synthetic trajectories based on Ornstein-Uhlenbeck (OU) process for confined motion and applying correlated random walk (CRW) model for ballistic motion, (step 2) evaluate trajectory properties, (step 3) quantify classification criteria for the statistical testing procedure, (step 4) import contact trajectories from experimental datasets, (step 5) evaluate trajectory properties like in step 2, and (step 6) classify trajectories based on the criteria quantified in step 3.} 
	\label{fig:methods_overview}
	\vspace{-0.3cm}
\end{figure*}

Our approach for contact interaction trajectory classification consists of two main parts: a statistical testing procedure (steps 1--3) and experimental dataset analysis (steps 4--6). For the statistical testing procedure, we first generate synthetic trajectories by simulating confined diffusion based on Ornstein-Uhlenbeck (OU) process and apply a correlated random walk (CRW) model for ballistic motion (step 1). In step 2, we then evaluate trajectory properties to reflect turning angle movement and the linearity of contact interaction trajectories. In step 3, we quantify classification criteria for the statistical testing procedure. For the experimental dataset analysis, we apply the statistical testing procedure to real data collected from pedestrian flow experiments. In doing so, we import contact trajectories from experimental datasets in step 4, and in step 5 we evaluate the trajectory properties as in step 2. In step 6, we then classify the trajectories based on the classification criteria that developed in step 3. A schematic representation of contact interaction trajectory classification can be seen from Fig.~\ref{fig:methods_overview}.

\subsection{Generation of synthetic trajectories}
\label{subsection:synthetic-trajectories}
We generate a set of synthetic contact interaction trajectories in line with random walk models. Two types of motions are considered for the trajectory generation: confined diffusion and ballistic motion. If a trajectory shows confined diffusion, the trajectory remains in a restricted area~\cite{Arinstein_PRE2005,Bickel_PhysicaA2007,CalvoMunoz_PRE2011}. In the context of contact interactions, confined diffusion results in long or even infinite duration of contact interaction among individuals. On the other hand, if a trajectory shows ballistic motion,  individuals keep their walking direction in the subsequent steps. Consequently, individuals tend to move in straight lines between their start and end points, likely showing short duration of contact interactions. We apply a stochastic differential equation (given in Eq.~(\ref{eq:sde_OU})) to generate the confined motion trajectories and correlated random walk model for the ballistic motion trajectories.

\begin{table}
	\normalsize                       %
	\setlength{\tabcolsep}{8pt}       % general space between columns (6pt standard) 
	\renewcommand{\arraystretch}{1.2} % general space between rows (1 standard)
	\centering
	\caption{Parameter setup for synthetic trajectory generation.}
	\label{table:synthetic_trajectory_generation_setup}
	\resizebox{8.5cm}{!}{
		\begin{tabular}{r*{3}{r}}
			\hline
			\hline			
			Motion type & Model & Parameter\\			
			\hline
			\hline			
			\multirow{4}{*}{Confined motion} & \multirow{4}{*}{OU} & $n = [10, 1000]$\\
			& & $\lambda = (\lambda_c, 1]$ with $\lambda_c = 0.1$\\
			& & $\mu = 0$ \\
			& & $\sigma = 1$ \\
			\hline
			\multirow{5}{*}{Ballistic motion} & \multirow{5}{*}{CRW} & $n = [10, 1000]$\\
			& & $\kappa = [0, 500]$\\
			& & $\beta_0 = 0$, $\beta(0) = 0$ \\			
			& & $v = 1$ \\
			& & $\Delta t = 1$ \\
			\hline			
			\hline
		\end{tabular}
	}
	\vspace{-0.0cm}
\end{table}

\subsubsection{Ornstein-Uhlenbeck (OU) process}
\label{subsubsubsection:OU-process}
According to Refs.~\cite{Briane_PRE2018,Weron_PRE2019,Janczura_2022}, we numerically simulate confined motion based on Ornstein-Uhlenbeck (OU) process, which is given in terms of the following stochastic differential equation:
\begin{equation} \label{eq:sde_OU}
dX_t = -\lambda (X_t-\mu)dt+ \sigma dB^{1/2}_t, 
\end{equation}
where $X_t = (x(t), y(t))$ is the position in the two-dimensional space obtained at time $t$. The equilibrium position is given as $\mu = (\mu_x, \mu_y)$. The interaction strength is indicated as $\lambda = (\lambda_x, \lambda_y)$, and its value follows a uniform distribution. Based on the parameter values presented in Table~\ref{table:synthetic_trajectory_generation_setup}, we simulate 10000 trajectories of the Ornstein-Uhlenbeck (OU) process in Eq.~(\ref{eq:sde_OU}). The simulation time step is set to 1 and the speed is controlled by means of the interaction strength $\lambda$. The simulation parameter values are selected in line with previous studies~\cite{Briane_PRE2018,Weron_PRE2019,Janczura_2022}.

\subsubsection{Correlated random walk (CRW) model}
\label{subsubsection:CRW}
We simulate ballistic motion based on the correlated random walk (CRW) model. For each simulation time step, we update the position of contact interaction trajectory for a given speed $v$:
\begin{align}
\label{eq:CRW}
\begin{split}
x(t+\Delta t) &= x(t) + v cos(\beta(t)) \Delta t,\\
y(t+\Delta t) &= y(t) + v sin(\beta(t)) \Delta t.\\
\end{split}
\end{align}
Here, $\Delta t$ is the time step size and $\beta(t)$ is the turning angle which is given as a
\begin{equation} \label{eq:tunring_angle}
	\beta(t) = \beta(t-\Delta t)+\Delta \beta. 
\end{equation}
In accordance with previous studies~\cite{Rutten_SciRep2022,Codling_2008,Liu_JTB2015,Fofana_ProcB2017}, the turning angle increment $\Delta \beta$ is sampled from the von Mises distribution $\nu(\beta_0|\kappa)$, where $\beta_0$ is the mean angle and $\kappa$ is the scale parameter reflecting directional correlation. Note that ballistic motion trajectories can be generated when the value of $\kappa$ is large. In contrast, if $\kappa$ is small, the generated trajectory tends to show pure random walk behavior. For each $\kappa$, we generate synthetic trajectories of ballistic motions by performing 10000 independent simulation runs. The time step $\Delta t$, is set as 1 and the speed $v$ is 1, so the walker moves a unit length at each time step.
The simulation parameter values in Table~\ref{table:synthetic_trajectory_generation_setup} are selected based on Refs.~\cite{Rutten_SciRep2022,Liu_JTB2015}.

\subsection{Trajectory characterization}
\label{subsection:trajectory_characterization}
We characterize contact interaction trajectories based on two test statistics: turning angle entropy $H$ and efficiency $E$. 

\subsubsection{Turning angle entropy}
\label{subsubsection:entropy}

\begin{figure}
	\centering
	\includegraphics[width=8.5cm]{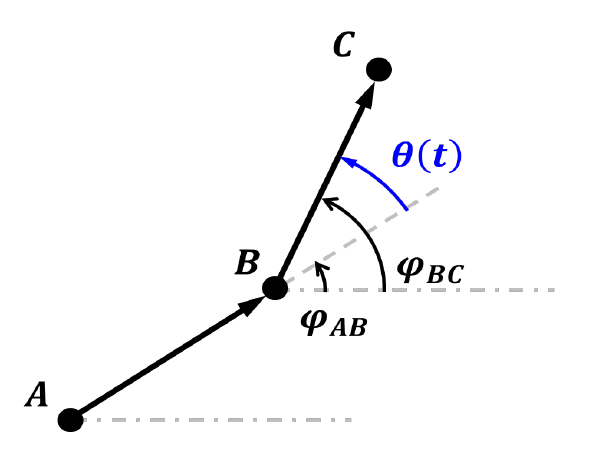}
	\caption{Schematic representation of turning angle. We compute the turning angle $\theta(t)$ at $B$ based on three successive positions $A$, $B$, and $C$, which represent the positions $[x(t-\tau), y(t-\tau)]$, $[x(t), y(t)]$, and $[x(t+\tau), y(t+\tau)]$, respectively. Here, $\varphi_{AB}$ is the angle of vector $\protect\overrightarrow{AB}$ relative to positive $x$ axis, and $\varphi_{BC}$ for $\protect\overrightarrow{BC}$. Note that $\theta$ is considered positive in counterclockwise direction.} 
	\label{fig:turning_angle}
\end{figure}

To understand the change of movement direction in contact interaction trajectories, we firstly obtained the turning angle $\theta(t)$. Similar to previous studies~\cite{Burov_PNAS2013,Parisi_PRE2016,Liu_PRE2018}, we compute the turning angle $\theta(t)$ based on three successive positions $[x(t-\tau), y(t-\tau)]$, $[x(t), y(t)]$, and $[x(t+\tau), y(t+\tau)]$, which are denoted by $A$, $B$, and $C$ in Fig.~\ref{fig:turning_angle}. The sampling interval $\tau$ determines how often we sample data points from a contact interaction trajectory to evaluate $\theta(t)$.

The turning angle $\theta(t)$ indicates the change in walking direction from the previous movement $\protect\overrightarrow{AB}$ to the current movement $\protect\overrightarrow{BC}$, i.e., $\varphi_{AB}$ and $\varphi_{BC}$ respectively. We calculate the angle $\varphi_{AB}$ of vector $\protect\overrightarrow{AB}$ relative to positive $x$ axis:
\begin{equation} \label{eq:varphi_AB}
	\varphi_{AB} = k_{AB} \pi + arctan\left[ \frac{y(t)- y(t-\tau)}{x(t)- x(t-\tau)} \right],
\end{equation}
with
\begin{equation*}
	k_{AB} =  
	\begin{cases} 
		0, &\text{if $x(t) > x(t-\tau)$ and $y(t) > y(t-\tau)$} \\
		1, &\text{if $x(t) < x(t-\tau)$} \\
		2, &\text{if $x(t) > x(t-\tau)$ and $y(t) < y(t-\tau)$}.\\
	\end{cases}
\end{equation*}

Likewise, we calculate the angle $\varphi_{BC}$ of vector $\protect\overrightarrow{BC}$ relative to positive $x$ axis:
\begin{equation} \label{eq:varphi_BC}
	\varphi_{BC} = k_{BC} \pi + arctan\left[ \frac{y(t+\tau)- y(t)}{x(t+\tau)- x(t)} \right],
\end{equation}
with
\begin{equation*}
	k_{BC} =  
	\begin{cases} 
		0, &\text{if $x(t+\tau) > x(t)$ and $y(t+\tau) > y(t-\tau)$} \\
		1, &\text{if $x(t+\tau) < x(t)$} \\
		2, &\text{if $x(t+\tau) > x(t)$ and $y(t+\tau) < y(t)$}.\\
	\end{cases}
\end{equation*}

Based on Equations~\ref{eq:varphi_AB} and~\ref{eq:varphi_BC}, the turning angle $\theta(t)$ is given as
\begin{equation} \label{eq:turning_angle}
	\theta(t) = 2m\pi+\varphi_{BC}-\varphi_{AB},
\end{equation}
with
\begin{equation*}
	m =  
	\begin{cases} 
		0, &\text{if $\left| \varphi_{BC}-\varphi_{AB} \right| < \pi$} \\
		-1, &\text{if $\left| \varphi_{BC}-\varphi_{AB} \right| > \pi$ and $\varphi_{BC} > \varphi_{AB}$}\\
		1, &\text{if $\left| \varphi_{BC}-\varphi_{AB}  \right| < \pi$ and $\varphi_{BC} < \varphi_{AB}$}.\\
	\end{cases}
\end{equation*}
Note that the range of turning angle $\theta(t)$ is between $-\pi$ and $\pi$.

For each contact interaction, we then define Shannon entropy of turning angle $H$ as
\begin{equation} \label{eq:entropy}
	H = -\sum_{i}^{N} p_i(\theta) \log_N p_i(\theta),
\end{equation}
where $N$ is the number of divided bins in the complete range of turning angle and $p_i(\theta)$ is the probability that a turning angle $\theta$ is observed in bin $i$. In this study, the complete range of turning angles $\theta \in [-\pi, \pi]$ are evenly divided into $N = 24$ sections with a constant size of $\Delta \theta = \pi/12 = 15^{\circ}$. We compute the probability $p_i(\theta)$ by counting the frequency of $\theta$ within each bin. The turning angle entropy $H$ lies between 0 and 1. When the contact interaction trajectory is a straight line, the turning angle entropy $H$ is 0, indicating that the turning angle is not changing during the contact interaction. In case of $H = 1$, the distribution of turning angle is uniform, indicating that every angle (bin) is observed, thus we can see that the turning angle is changing frequently. 

\subsubsection{Efficiency}
\label{subsubsection:efficiency}
Efficiency $E$ measures the ratio of the net squared displacement of a trajectory to the sum of squared step lengths, reflecting the linearity of the trajectory. For a trajectory containing $n$ data points of position, Efficiency $E$ is defined as 
\begin{equation} \label{eq:efficiency}
	E = \frac{|X_{n-1}-X_0|^2}{\sum_{i = 1}^{n-1} |X_{i}-X_{i-1}|^2},
\end{equation}
where $X_{i}$ denotes the position at time instance $i$ and $X_{0}$ for the position at the start of contact interaction.

\subsection{Classification criteria}
\label{subsection:classification_criteria}
In line with the statistical test procedure presented in previous studies~\cite{Briane_PRE2018,Janczura_PRE2020,Weron_PRE2019}, we identify classification criteria for different motion types based on the knowledge of turning angle entropy $H$ and efficiency $E$ distributions that we can observe from the synthetic trajectories. 

\begin{figure}
	\centering\includegraphics[width=6.5cm]{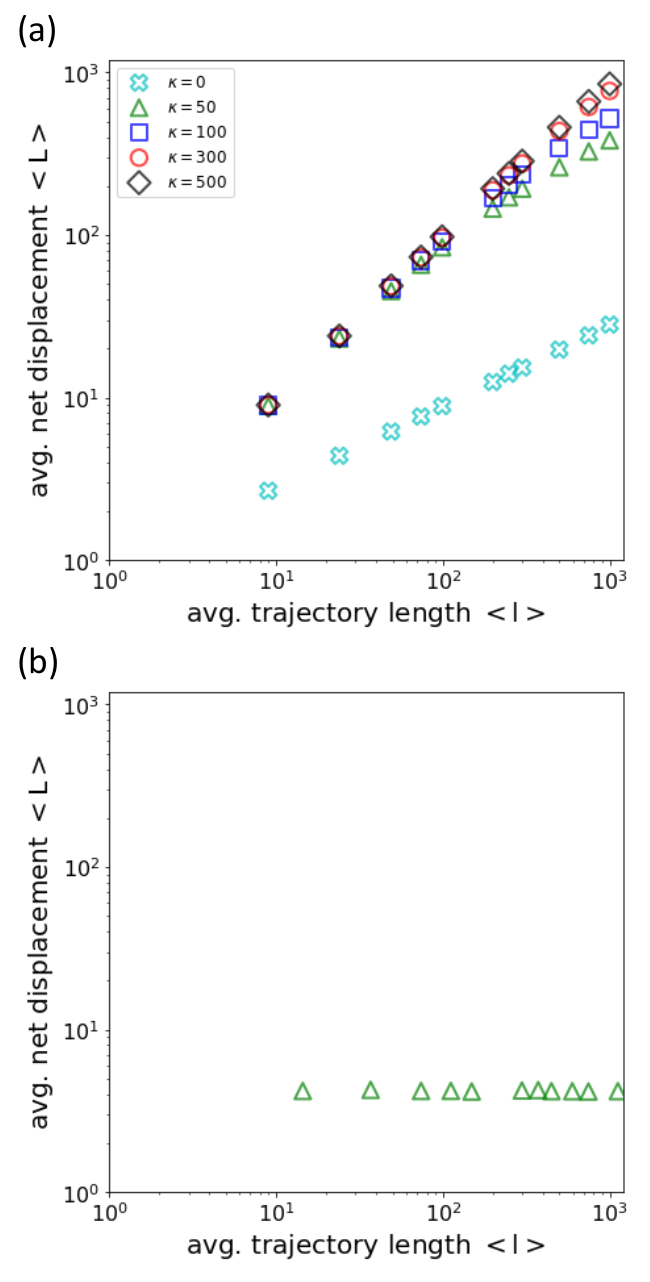}
	\vspace{-0.0cm}
	\caption{The average end-to-end distance $\left\langle L \right\rangle$ as a function of average trajectory length $\left\langle l \right\rangle$ measured from: (a) correlated random walk (CRW) trajectories and (b) Ornstein-Uhlenbeck (OU) process trajectories.} 
	\label{fig:scaling_behavior}
\end{figure}

First, we evaluate the scaling behavior of average end-to-end distance $\left\langle L \right\rangle$ as a function of average trajectory length $\left\langle l \right\rangle$. By doing that, we can quantify the confined and ballistic motions from OU process and CRW model simulations, respectively. Previous studies~\cite{Huang_PRE2002,Visser_2006} suggested that one can classify a random walk motion as a ballistic motion if the slope of $\left\langle L \right\rangle$ the curve in a log–log plot is approximately equal to 1, i.e., $\left\langle L \right\rangle \sim \left\langle l \right\rangle$. In addition, it is also suggested that $\left\langle L \right\rangle$ the curve reaches a plateau (i.e., the slope of $\log(\left\langle L \right\rangle)$ curve is near 0) in the case of confined motion. That is, the trajectory of contact interaction is trapped in a small area, thus the end-to-end distance $L$ does not increase further, even though the trajectory length $l$ increases. Figure~\ref{fig:scaling_behavior} shows a log–log plot of average end-to-end distance $L$ as a function of average trajectory length $l$ for the CRW model and OU process simulations. Analogous to previous studies~\cite{Janczura_PRE2020,Weron_PRE2019}, we estimate the slope of $\left\langle L \right\rangle$ curve by fitting a function $a\log(t)+b$ to the estimated $\log(\left\langle L \right\rangle)$ curve. We choose the cutoff value of the exponent as $c = 0.05$. When the slope of $\left\langle L \right\rangle$ curve in a log–log plot is between $1-c = 1.95$ and $1+c = 2.05$, i.e., $a \in [1.95, 2.05]$, we consider the corresponding trajectory shows ballistic motion. If the value of $a$ is between $-c = -0.05$ and $c = 0.05$, i.e., $a \in [-0.05, 0.05]$, the corresponding trajectory is classified as confined motion. In line with the scaling behavior of $\left\langle L \right\rangle$ presented in Fig.~\ref{fig:scaling_behavior}, we categorize synthetic trajectories as the ballistic motion if the trajectories are generated by Eq.~(\ref{eq:CRW}) with $\kappa \geq 300$, and trajectories generated by Eq.~(\ref{eq:sde_OU}) as the confined motion.

\begin{figure}
	\centering\includegraphics[width=8.0cm]{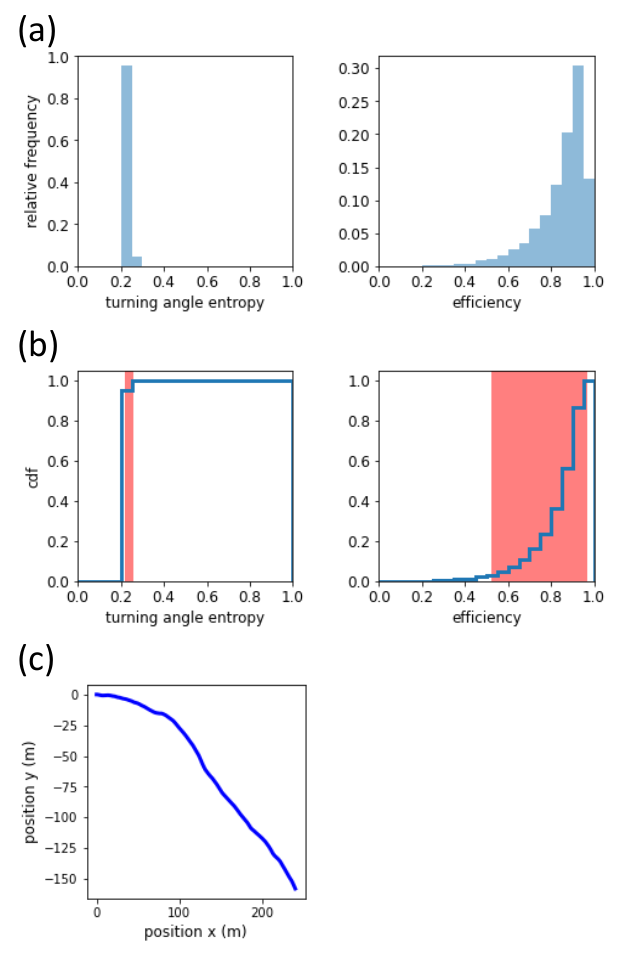}
	\vspace{-0.0cm}
	\caption{A representative example of turning angle entropy $H$ and efficiency $E$ distributions for the ballistic motion: (a) relative frequency and (b) cumulative distribution. A sample trajectory is shown in (c). The red shaded area in (b) indicates the critical region with the significance level $\alpha = 0.05$ suggested in the previous studies Refs.~\cite{Briane_2016,Briane_PRE2018,Weron_PRE2019}. Distribution curves are generated based on 10000 independent simulation runs with $\kappa = 300$ and $n = 300$.} 
	\label{fig:indicator_dist_ballistic}
\end{figure}

\begin{figure}
	\centering\includegraphics[width=8.0cm]{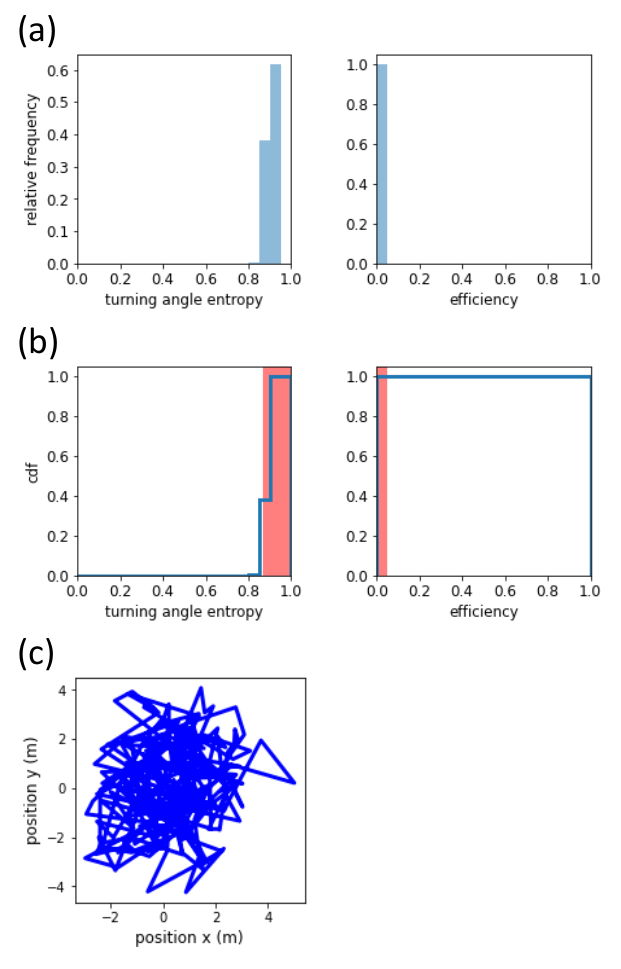}
	\vspace{-0.0cm}
	\caption{Same as Fig.~\ref{fig:indicator_dist_ballistic} but for the confined motion. Distribution curves are generated based on 10000 independent simulation runs with $n = 300$.} 
	\label{fig:indicator_dist_confined}
\end{figure}

Next, we define a critical region of turning angle entropy $H$ and efficiency $E$ distributions for the ballistic and confined motions. For different numbers of (positional) data points, we estimate the boundaries of the critical region for test statistics $H$ and $E$. For the simulated trajectories generated by the CRW model with $\kappa \geq 300$, we assume that ballistic motion will tend to show low $H$ and high $E$. Figure~\ref{fig:indicator_dist_ballistic} shows a representative example of turning angle entropy $H$ and efficiency $E$ distributions for ballistic motion. As can be seen from Table~\ref{table:criteria_ballistic}, the upper boundary of $H$ is identified as
\begin{equation} \label{eq:ballistic_upper_H}
	H \leq q_{1-\alpha},
\end{equation}
where $q_{1-\alpha}$ indicates that $H$ lies in the critical region with the probability $1-\alpha$. The lower boundary of $E$ is identified as 
\begin{equation} \label{eq:ballistic_lower_E}
	q_{\alpha} \leq E,
\end{equation}
where $q_{\alpha}$ indicates that $E$ lies in the critical region with the probability $1-\alpha$. We use $\alpha = 0.05$ according to Refs.~\cite{Briane_PRE2018,Weron_PRE2019}. One can notice that the turning angle entropy $H$ is stable, while the efficiency $E$ is notably decreasing as the number of data points $n$ increases. During the ballistic motion, individuals tend maintain their walking direction in the subsequent steps, thus the change in walking direction between two consecutive steps is small, yielding low $H$. As the individuals continue walking (i.e., $n$ is increasing), sometimes the change in walking direction adds up, resulting in deviation from a straight line and consequently leading to decreasing $E$ for increasing $n$. Thus, it is suggested that the turning angle entropy $H$ is a useful measure for the ballistic motion. 

In contrast to the ballistic motion trajectories, the confined motion trajectories generated using the OU process yield high $H$ and low $E$ as shown in Fig.~\ref{fig:indicator_dist_confined}. Similarly to Eqs.~(\ref{eq:ballistic_upper_H}) and (\ref{eq:ballistic_lower_E}), we identify the lower boundary of $H$ as
\begin{equation} \label{eq:confined_lower_H}
	q_{\alpha} \leq H,
\end{equation}
and the upper boundary of $E$ as
\begin{equation} \label{eq:confined_upper_E}
	E \leq q_{1-\alpha}.
\end{equation}
In Table~\ref{table:criteria_confined}, the turning angle entropy $H$ is increasing rapidly for the growing number of data points $n$ particularly for $n < 100$, but the efficiency $E$ is virtually constant for a wide range of $n$. During the confined motion, individuals continue changing walking direction and the number of data points $n$ is increasing. When $n$ is small, the turning angle entropy $H$ might be underestimated due to the limited number of data points. In this case, the value of $H$ is increasing as $n$ grows, which can be understood that $H$ is reaching the stationary state value once $n$ becomes sufficiently large. Meanwhile, $E$ is stable for a wide range of $n$, suggesting that the efficiency $E$ is a good measure for the confined motion.   
 
\begin{table}
	\normalsize                       %
	\setlength{\tabcolsep}{6pt}       % general space between columns (6pt standard) 
	\renewcommand{\arraystretch}{1.2} % general space between rows (1 standard)
	\centering
	\caption{Classification criteria for ballistic motion. For different number of data points $n$, we evaluate the critical region in the distributions of turning angle entropy $H$ and efficiency $E$ with the significance level $\alpha = 0.05$.}
	\label{table:criteria_ballistic}
	\resizebox{8.5cm}{!}{
		\begin{tabular}{rrcccccc}
			\hline
			\hline
			& & \multicolumn{6}{c}{number of data points $n$}\\
			\cline{3-8}
			test statistics & & 10 & 25 & 50 & 100 & 500 & 1000\\	
			\hline
			entropy $H$ & smaller than & 0.21 & 0.22 & 0.22 & 0.22 & 0.26 & 0.26\\
			efficiency $E$ & larger than & 0.98 & 0.96 & 0.91 & 0.82 & 0.33 & 0.11\\
			\hline
			\hline
		\end{tabular}
	}
	\vspace{-0.0cm}
\end{table}

\begin{table}
	\normalsize                       %
	\setlength{\tabcolsep}{6pt}       % general space between columns (6pt standard) 
	\renewcommand{\arraystretch}{1.2} % general space between rows (1 standard)
	\centering
	\caption{Same as Table~\ref{table:criteria_ballistic} but for the confined motion.}
	\label{table:criteria_confined}
	\resizebox{8.5cm}{!}{
		\begin{tabular}{rrcccccc}
			\hline
			\hline
			& & \multicolumn{6}{c}{number of data points $n$}\\
			\cline{3-8}
			test statistics & & 10 & 25 & 50 & 100 & 500 & 1000\\			
			\hline
			entropy $H$ & larger than & 0.21 & 0.53 & 0.68 & 0.78 & 0.89 & 0.90\\
			efficiency $E$ & smaller than & 0.09 & 0.05 & 0.05 & 0.05 & 0.05 & 0.05\\
			\hline
			\hline
		\end{tabular}
	}
	\vspace{-0.0cm}
\end{table}

\begin{figure}
	\centering\includegraphics[width=6.5cm]{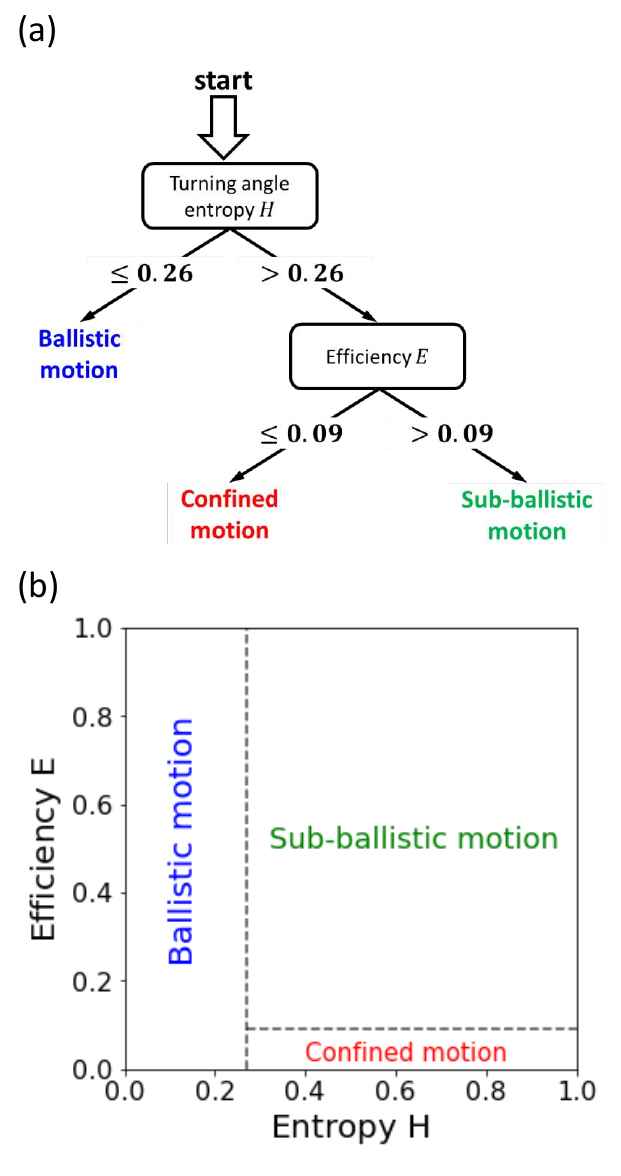}
	\vspace{-0.0cm}
	\caption{Summary of classification criteria: (a) a classification tree  (b) a schematic diagram presenting the criteria in $(H, E)$ space.} 
	\label{fig:classification_criteria_summary}
\end{figure}

To identify classification criteria applicable over a wide range of $n$, we use turning angle entropy $H$ and efficiency $E$ to characterize the ballistic and confined motions, respectively. Consequently, the ballistic motion is characterized by 
\begin{equation} \label{eq:characterize_ballistic}
	H \leq 0.26,
\end{equation}
and the confined motion is characterized by 
\begin{equation} \label{eq:characterize_confined}
	E \leq 0.09.
\end{equation}
In addition, we define sub-ballistic motion for the remainder of the parameter space, which does not belong to either ballistic or confined motions. Figure~\ref{fig:classification_criteria_summary} summarizes the presented classification criteria. It is noted that change in the cutoff value $c$ affects the threshold values of $H$ and $E$ shown in Eqs.~(\ref{eq:characterize_ballistic})~and~(\ref{eq:characterize_confined}), but the scaling behavior of average end-to-end distance shown in Fig.~\ref{fig:scaling_behavior} does not directly influence on the computation of $H$ and $E$ shown in Eqs.~(\ref{eq:entropy})~and~(\ref{eq:efficiency}).
%% thus the boundaries of ballistic and confined motions in $(H, E)$ space in Figure~\ref{fig:classification_criteria_summary} shift.

\section{Results and Discussion}
\label{section:results}

\subsection{Datasets}
\label{subsection:datasets}

\begin{figure*}
	\centering\includegraphics[width=16cm]{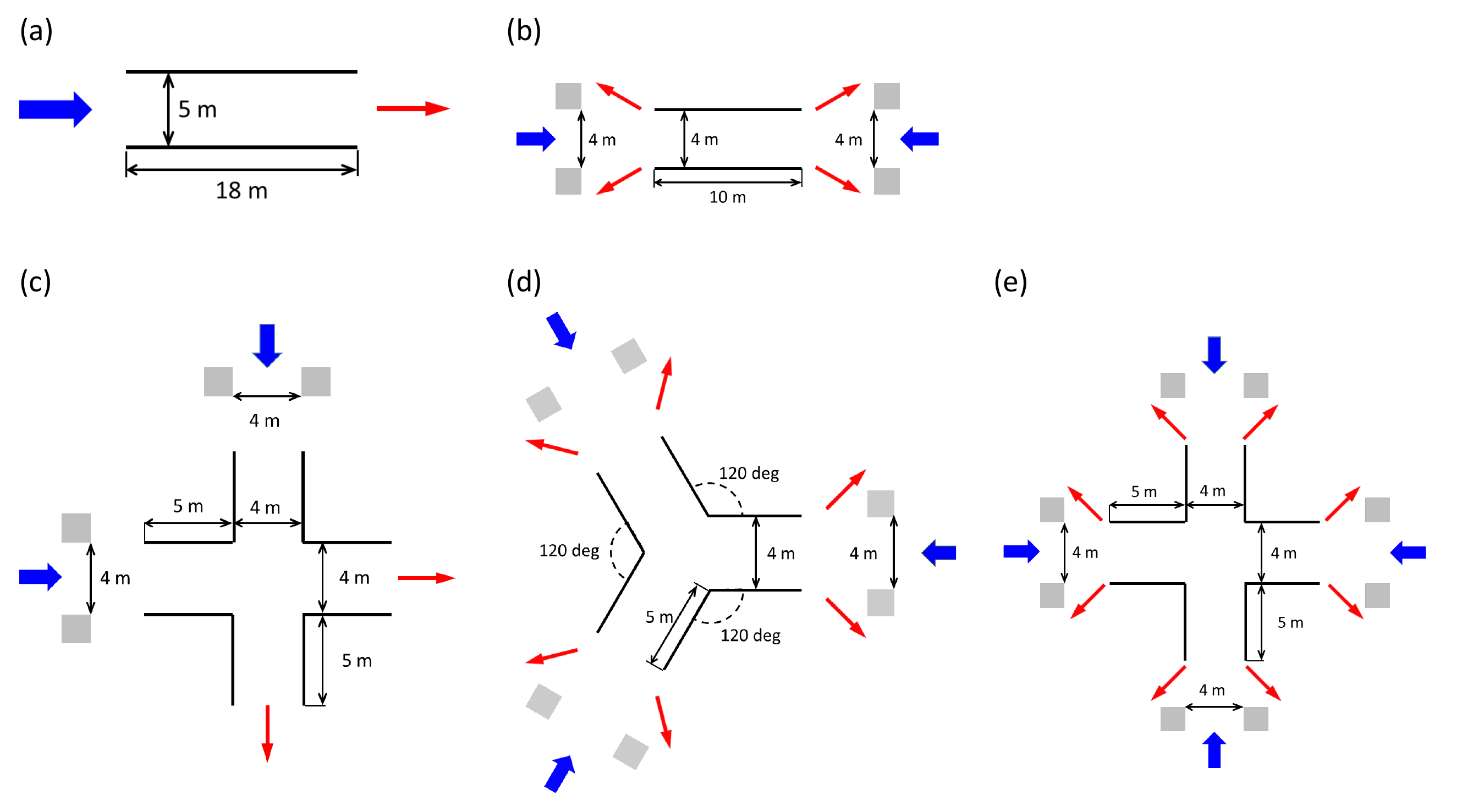}
	\vspace{-0.0cm}
	\caption{Schematic representation of experiment setups: (a) unidirectional flow (scenario name: uni-05), (b) bidirectional flow (scenario name: bi-b01), (c) 2-way crossing flow (scenario name: crossing-90-d08), (d) 3-way crossing flow (scenario name: crossing-120-b01), and (e) 4-way crossing flow (scenario name: crossing-90-a10). Here, blue thick arrows show the walking direction of incoming pedestrians entering the corridors and red thin arrows indicates the walking direction of outgoing pedestrians leaving the corridors. A pair of gray rectangles is placed to set up an entrance of pedestrian group entering the corridor.} 
	\label{fig:experiment_setup}
\end{figure*}

Figure~\ref{fig:experiment_setup} shows the sketches of the various experimental setups: uni-directional flow, bi-directional flow, 2-way crossing flow, 3-way crossing flow, and 4-way crossing flow. In the uni-directional flow setup, pedestrians were walking to the right in a straight corridor of 5~m wide and 18~m long. In the bi-directional flow case, two groups of pedestrians enter a straight corridor of 4~m wide and 10~m long through 4~m wide entrance and then walking opposite directions. They leave the corridor through the open passage once they reached the other side of the corridor. In the 2-way, 3-way, and 4-way crossing flows, different groups of pedestrians enter the corridor through 4~m wide entrance and walked 5~m before and after passing through an intersection (4~m by 4~m rectangle in 2-way crossing and 4-way crossing flows, and 4~m wide equilateral triangle in 3-way crossing flow). Similar to the setup of the bi-directional flow, pedestrian groups leave the corridor through the open passage at the end of the corridors. Trajectories of the bi-directional, 2-way, and 3-way crossing flows were recorded at 16 frames per second (fps) and 25 fps for uni-directional and 4-way crossing flow. A more detailed description of the experiment setups can be found in Refs.~\cite{Holl_Dissertation2016,Cao_JSTAT2017,url_dataset}.

From the experimental data, we extracted pairs of individuals in contact and their relative motion trajectories. We consider a pair of individuals to be in contact when the two individuals are within a contact radius. The contact radius $r_c$ would depend on the form of transmission in question, and is thus relevant to the evaluation of direct transmission risk. We have chosen a 2m radius based on existing literature~\cite{Rutten_SciRep2022,Han_Lancet2020,Ronchi_SafetySci2020}, so that we can capture a wide range of common infectious diseases. It is also important to note that while we focus on spatial proximity as a measure of risk, there are many other factors that influence the risk of transmission (e.g., head orientations, respiratory activities, and airflow conditions) and that is highly disease dependent~\cite{Garcia_SafetySci2021,Mendez_AdvSci2023,Nicolas_2023,Rahn_plos2022,Bale_SciRep2022}. For the analysis, we consider contact interaction trajectories with at least 0.5 seconds records, i.e., 8 data points for the bi-directional, 2-way, and 3-way crossing flows, and 12 data points for the uni-directional and 4-way crossing flow based on literature~\cite{Briane_PRE2018,Weron_PRE2019}. Table~\ref{table:basic_statistics} shows the basic statistics of the selected scenarios, including the number of individuals $N$, experiment period, and the number of contacts. In this study, we consider three types of contact interactions: parallel, head-on, and crossing contact interactions. When two individuals in contact are moving in the same direction, one can define the parallel contact interaction. A head-on contact interaction is observed when two individuals are moving toward opposite directions. We can see a crossing contact interaction from a pair of individuals in contact when one individual is moving perpendicular to the other's moving direction. The number of contacts is given as the number of interacting individual pairs. In Appendix~\ref{sec:appendix_basic_statistics}, we present the basic descriptive statistics of the other experimental scenarios.
%% Although the use of physical proximity is a popular choice for evaluating the risk of direct transmissions, the transmission risks might not be very sensitive to the inter-personal distance~\cite{Garcia_SafetySci2021,Mendez_AdvSci2023,Nicolas_2023,Rahn_plos2022}. 

\begin{table*}
	\normalsize                       %
	\setlength{\tabcolsep}{6pt}       % general space between columns (6pt standard) 
	\renewcommand{\arraystretch}{1.2} % general space between rows (1 standard)
	\centering
	\caption{Basic descriptive statistics of representative scenarios.}
	\label{table:basic_statistics}
	\resizebox{15.5cm}{!}{
		\begin{tabular}{c*{9}{r}}
			\hline
			\hline
			& & & & & \multicolumn{4}{c}{No. contacts}\\
			\cline{6-9}
			Setup & fps & Scenario name & N & Period (s) & Parallel & Head-on & Crossing & total\\
			\hline
			Uni-directional & 25 & uni-05 & 905 & 157.68 & 25260 (100\%) & 0 & 0 & 25260\\
			Bi-directional & 16 & bi-b10 & 736 & 324.56 & 17154 (29.81\%) & 40398 (70.19\%) & 0 & 57552\\
			2-way crossing & 16 & crossing-90-d08 & 592 & 147.88 & 22418 (53.39\%) & 0 & 19574 (46.61\%) & 41992\\
			3-way crossing & 16 & crossing-120-b01 & 769 & 215.63 & 34654 (36.80\%) & 59518 (63.20\%) & 0 & 94172\\
			4-way crossing & 25 & crossing-90-a10 & 324 & 94.08 & 882 (9.60\%) & 3834 (41.73\%) & 4472 (48.67\%) & 9188\\
			\hline
			\hline
		\end{tabular}		
	}
	\vspace{-0.0cm}
\end{table*}

\subsection{Classification results}
\label{subsection:classification-results}

\begin{figure*}
	\centering
	\includegraphics[width=14.0cm]{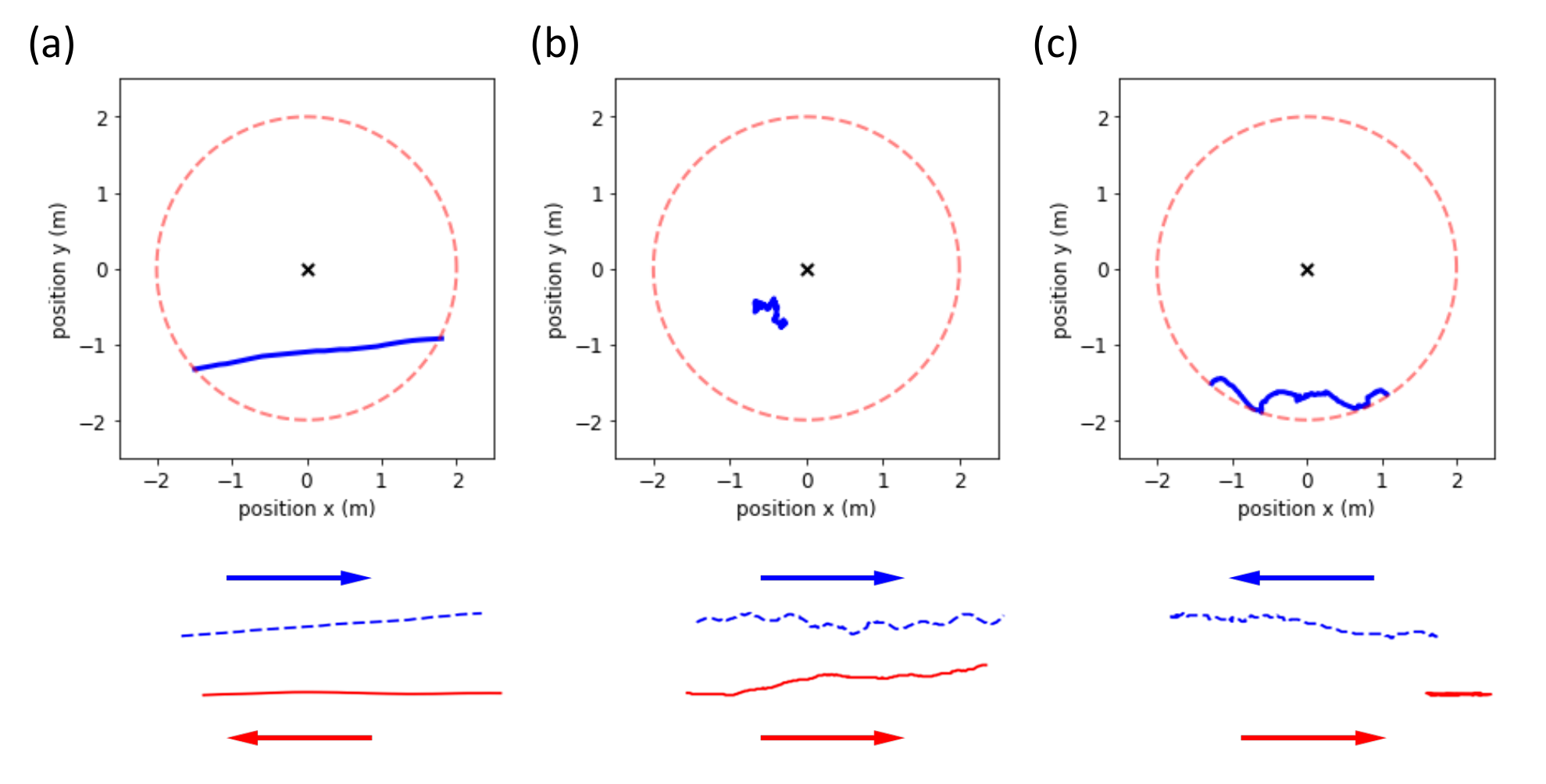}
	\vspace*{-2mm}
	\caption{Representative types of pedestrian contact interaction trajectories characterized based on the value of turning angle entropy $H$ and efficiency $E$ (refer to  Eqs.~(\ref{eq:characterize_ballistic}) and (\ref{eq:characterize_confined}), and Fig.~\ref{fig:classification_criteria_summary}): (a) ballistic motion with $H = 0.21$ and $E = 0.99$ (individual id 65 and id 99 in 4-way crossing scenario crossing-90-a10), (b) confined motion with $H = 0.87$ and $E = 0.02$ (individual id 25 and id 46 in unidirectional flow scenario uni-05), and (c) sub-ballistic motion with $H = 0.71$ and $E = 0.38$ (individual id 463 and id 566 in bi-directional scenario bi-b10). The position of focal individuals (id 25 in (a), id 65 in (b), and id 463 in (c)) is indicated at (0, 0) by a black cross symbol $\times$. Red dashed circles show contact range of a focal individual ($r_c = 2$~m). The relative motion of pedestrians interacting with the focal individuals (id 46 in (a), id 99 in (b), and id 566 in (c)) denoted by blue solid lines. In the lower panels, blue dashed lines indicate the ground truth trajectories of focal individuals (id 65 shown in (a), id 25 in (d), and id 463 in (c)), and red solid lines for pedestrians interacting with the focal individuals (id 99 shown in (a), id 46 in (b), and id 566 in (c)). Arrows are guide for the eyes, indicating the walking direction of individuals.} 
	\label{fig:sample_trajectory}
\end{figure*}

Figure~\ref{fig:sample_trajectory} presents representative types of pedestrian contact interaction trajectories characterized by turning angle entropy $H$ and efficiency $E$ (see Eqs.~(\ref{eq:characterize_ballistic}) and (\ref{eq:characterize_confined}), and Fig.~\ref{fig:classification_criteria_summary})). As can be seen from Fig.~\ref{fig:sample_trajectory}(a), the relative motion of individual $j$ moves in parallel along a straight line, showing a ballistic motion. In Fig.~\ref{fig:sample_trajectory}(b), the relative motion trajectory of individual $j$ stays close to its initial position during the contact interaction, implying confined motion. It should be noted that confined motion in contact interaction trajectories does not necessarily suggest that ground truth trajectories display confined motions. In the case of sub-ballistic motion (see Fig.~\ref{fig:sample_trajectory}(c)), the relative motion of individual $j$ occasionally shows changes in its walking direction but still travels far away from its initial position during the contact interaction. 

\begin{figure*}
	\centering
	\includegraphics[width=15.5cm]{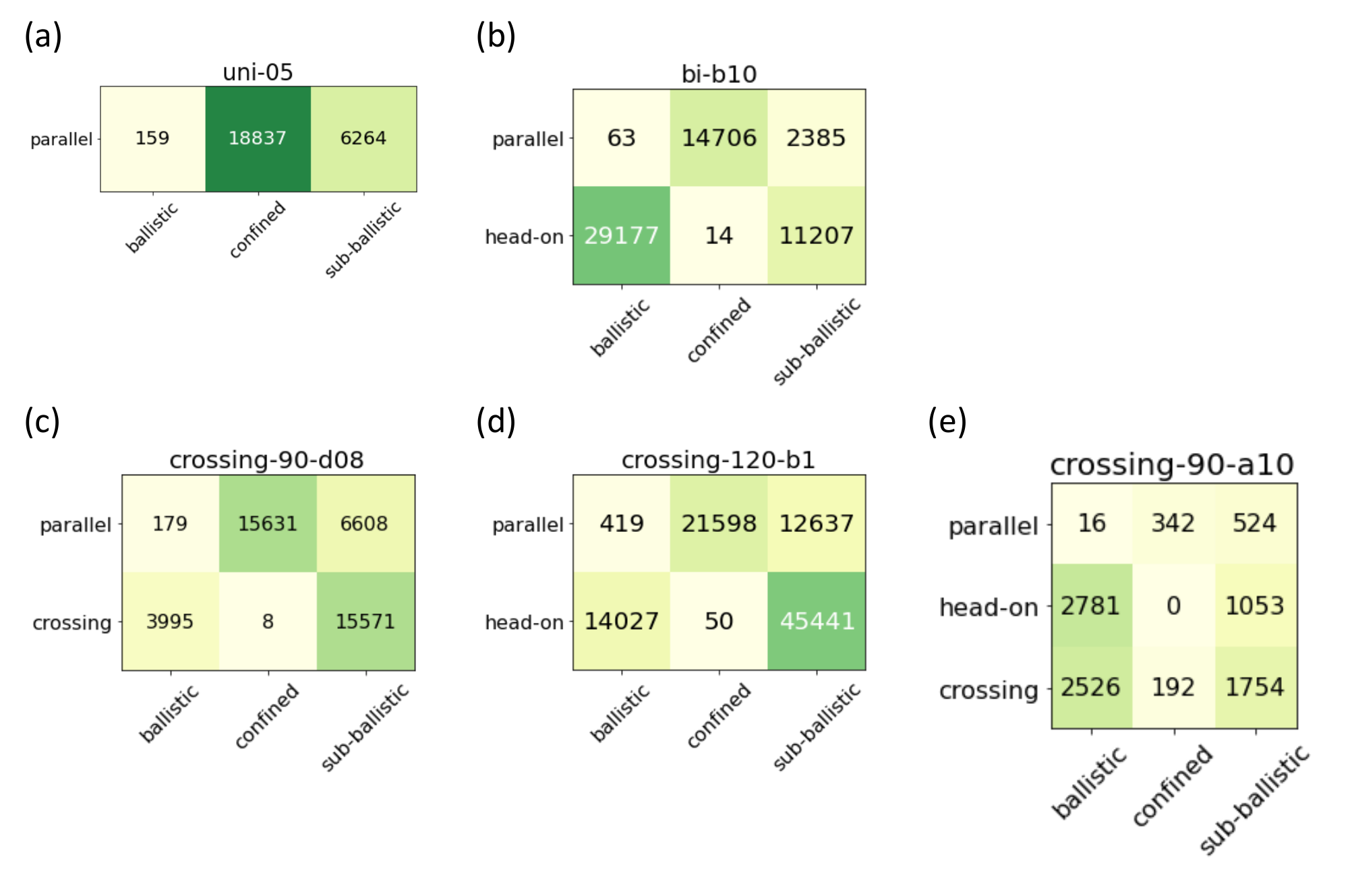}
	\caption{Summary of motion type classification results. For each panel, horizontal rows correspond to contact interaction types and vertical columns to the motion types.} 
	\label{fig:motion_type_classification}
\end{figure*}

Based on the classification criteria presented in section~\ref{subsection:classification_criteria}, we have classified motion types of contact interaction trajectories. The summary of the motion type classification results for the selected scenarios can be found in Fig.~\ref{fig:motion_type_classification}. We can observe that ballistic motion is more frequently observed in the presented scenarios of the bi-directional (scenario bi-b10) and 4-way crossing flow (scenario crossing-90-a10). In contrast, the majority of contact interactions in the unidirectional flow setup (scenario uni-05) are categorized as confined motion, hinting at the possibility of long-lived contact duration. Such long-lived contact durations provide a longer time in which transmission can occur between a pair of individuals. However, in the case where individuals experience many long-lived interactions, it necessarily means fewer total interactions (i.e., a few long-term interactions) because an individual may have only a limited number of neighbors within a contact radius. The risk of the transmission then depends on the infectiousness of the disease, with long-lived interactions providing opportunity for transmission even in low transmissibility cases. This is in line with the findings of previous studies~\cite{Rutten_SciRep2022,Smieszek_2009}, demonstrating that the infectious disease transmission risk would be considerable even for low transmissibility diseases. In addition, for other experiment scenarios including crossing-90-d08 (2-way crossing setup) and crossing-120-b01 (3-way crossing setup), the sub-ballistic motion is the most frequently observed motion type. A summary of motion type classification results for other experiment scenarios are given in Appendix~\ref{sec:appendix_classification_results}. 
%%% In line with the findings of previous studies~\cite{Rutten_SciRep2022,Smieszek_2009}, it can be suggested that the infectious disease transmission risk would be considerable even for low transmissibility. 
% low transmissibility --> low infectivity???

\begin{figure*}
	\centering
	\includegraphics[width=15.5cm]{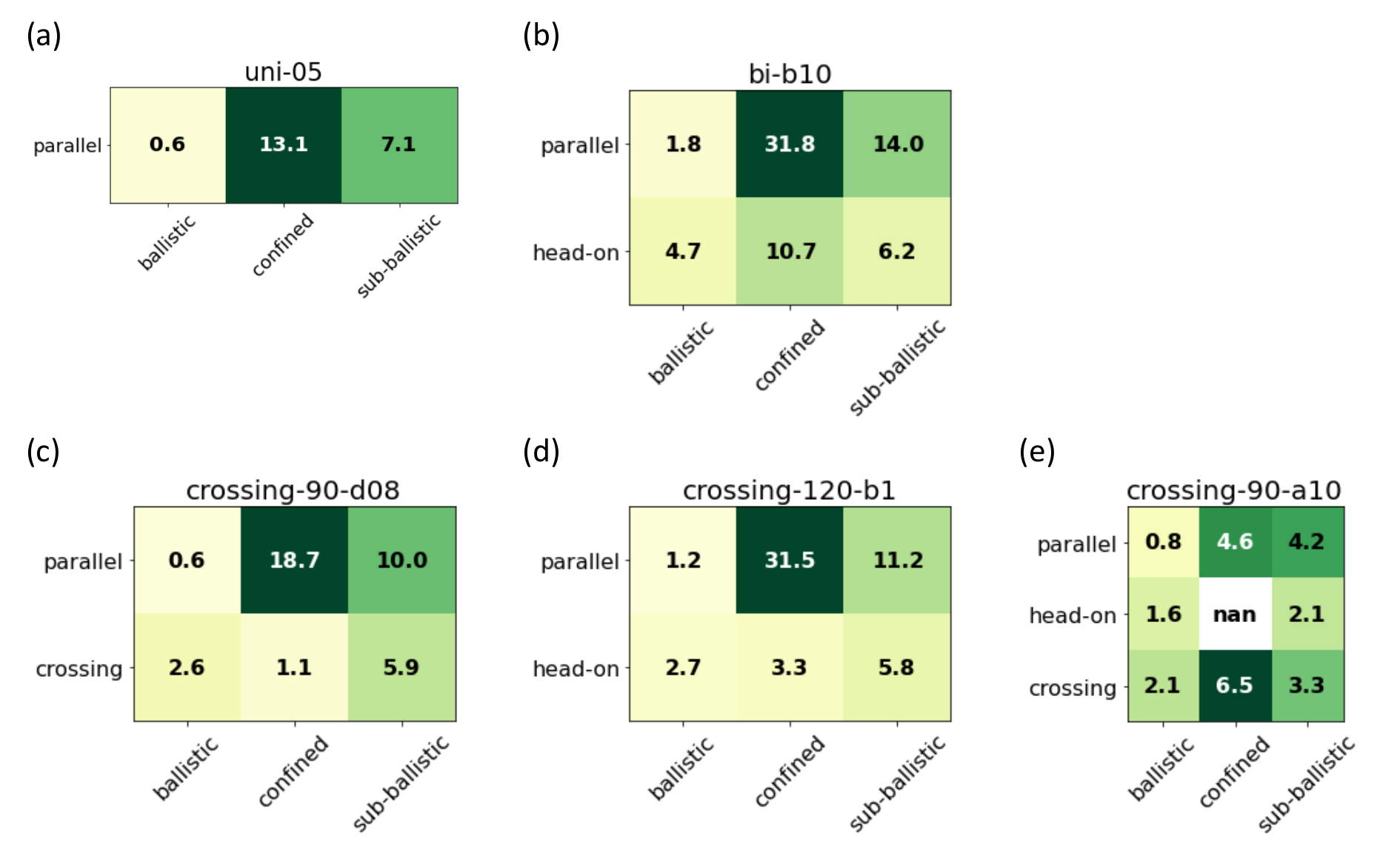}
	\caption{Average values of contact duration $t_c$ measured for different contact trajectory type classification. For each panel, horizontal rows correspond to contact interaction types and vertical columns to the motion types. Dark green cell indicates high value of $t_c$.} 
	\label{fig:motion_type_tc}
\end{figure*}

Figure~\ref{fig:motion_type_tc} illustrates average values of contact duration $t_c$ measured for different contact trajectory type classification. For the presented scenarios, confined motions observed from parallel contact interaction tend to yield high value $t_c$, displaying striking difference from $t_c$ measured from ballistic motions. This tendency can be also seen from other scenarios, refer to Appendix~\ref{sec:appendix_tc_avg}.

\begin{figure*}
	\centering
	\includegraphics[width=16.5cm]{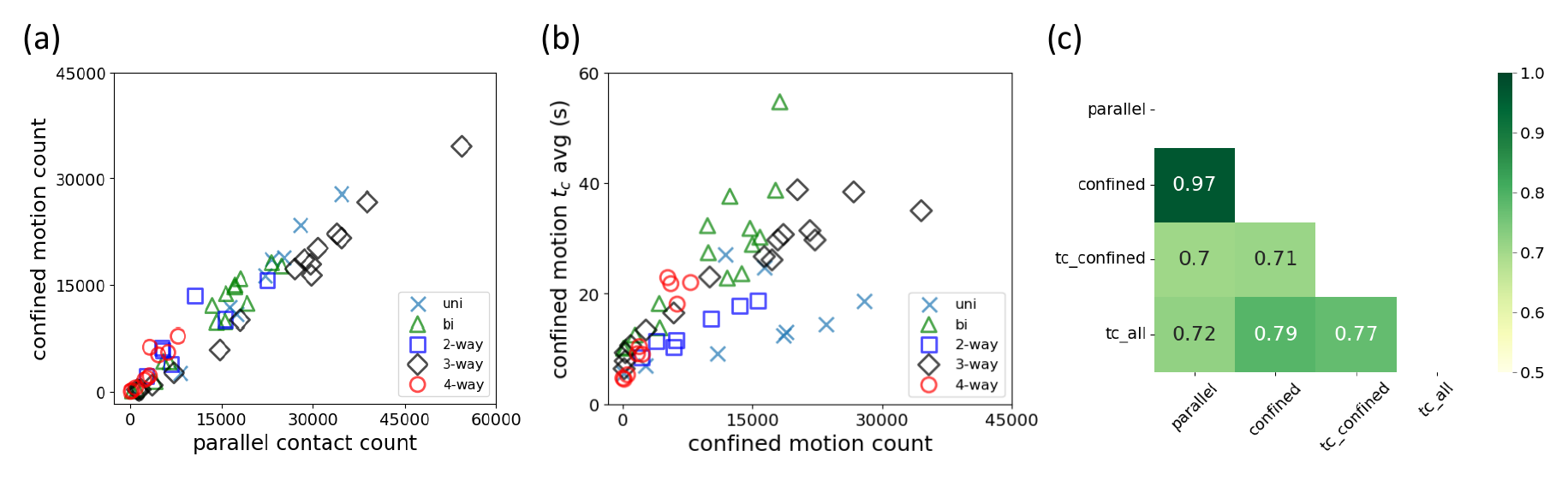}
	\caption{Relationship among the parallel contact interactions, confined motion, and contact duration $t_c$: (a) the frequency of confined motion increases as that of parallel contact interaction grows, (b) the average value of contact duration $t_c$ measured for the confined motions tends to increase as more confined motions are observed from the experiments, and (c) correlation matrix showing. In (a) and (b), different symbols indicate different experiment setup. In (c), "tc\_all" indicates the average contact duration evaluated for each experiment scenario regardless of motion types. The frequency of parallel contact interactions and confined motions are indicated by "parallel" and "confined", respectively.} 
	\label{fig:relationship_parallel_confined_tc}
\end{figure*}

To understand the influence of confined motions observed from parallel contact interaction on the contact duration $t_c$, we take a closer look at the relationship among the parallel contact interactions, confined motion, and $t_c$, see Fig.~\ref{fig:relationship_parallel_confined_tc}. As in shown Fig.~\ref{fig:relationship_parallel_confined_tc}(a), the frequency of confined motion increases as the frequency of parallel contact interaction grows. One can see from Fig.~\ref{fig:relationship_parallel_confined_tc}(b) that the average value of contact duration $t_c$ measured for the confined motions shows an increasing trend when the confined motions are more frequently observed from the experiments. Based on Fig.~\ref{fig:relationship_parallel_confined_tc}(a) and Fig.~\ref{fig:relationship_parallel_confined_tc}(b), it is suggested that more parallel contact interactions tend to produce more confined motions, and consequently yields longer contact duration $t_c$. That is compatible with high correlation among the parallel contact interactions, confined motion, and contact duration $t_c$ presented in Fig.~\ref{fig:relationship_parallel_confined_tc}(c).

\section{Conclusion}
\label{section:conclusion}
In this study, we have examined how the pedestrian flow setups affect contact duration distribution which is highly relevant to the spreading risk in human crowds. 
To this end, we have analyzed pedestrian contact interaction trajectories of uni-, bi- and multi-directional flow setups in experimental data~\cite{Holl_Dissertation2016,Cao_JSTAT2017,url_dataset}. In line with statistical testing approach, we have classified types of motions observed in the contact interactions based on the turning angle entropy $H$ and efficiency $E$. We generated synthetic trajectories and then evaluated the trajectory properties to systematically quantify classification criteria. The identified criteria were then applied to classify the contact interaction trajectories collected from pedestrian flow experiments. 

From the experimental dataset, we have classified the pedestrian contact interaction trajectories for three types of motions: ballistic, confined, and sub-ballistic motions. We observed the ballistic motion when $H$ is low. In the case of ballistic motion, the contact interaction trajectories move in parallel along a straight line, so the contact duration tends to be brief. On the other hand, the confined motion can be defined when $H$ is high and $E$ is low. This indicates that the contact interaction trajectories stay close to their initial position during the contact interaction. From the experimental data of various pedestrian flow setups, we have found that confined motion appears more as the frequency of parallel contact interactions increases. It is noted that the confined motions tend to yield longer contact duration $t_c$, potentially leading to higher risk of virus exposure in human face-to-face interactions. 

A few selected experimental setups have been analyzed to study the fundamental role of pedestrian flow setups (e.g., uni-, bi-, and multi-directional flow) in the distribution of pedestrian motion types and contact duration. To generalize the findings of this study, the presented analysis should be further performed with larger number of scenarios and various layouts of pedestrian facilities at different locations, for instance, mass gathering events like concerts and sport games. The presented approach can be applied to support facility managers and decision makers by quantifying the contact interaction characteristics for the crowd movement trajectories. Another interesting extension of the presented study can be planned in line with the importance analysis of factors influencing on the motion types and duration of contact interactions~\cite{Janczura_PRE2020,Kowalek_JPhysA2022,Kowalek_PRE2019,Wagner_PLOS2017,Pinholt_PNAS2021}.

\section*{Acknowledgements}
This research is supported by Ministry of Education (MOE) Singapore under its Academic Research Fund Tier 1 Program Grant No. RG12/21 MoE Tier 1 and by the Singapore Ministry of Health’s National Medical Research Council under its National Epidemic Preparedness and Response R\&D Funding Initiative (MOH-001041) Programme for Research in Epidemic Preparedness And REsponse (PREPARE).

% \newpage\newpage
\clearpage\clearpage
\appendix

\section{Data and code availability}
\label{section:data-and-code}
The data used for our study can be found from \url{https://ped.fz-juelich.de/da/doku.php?id=start#data_section} (last accessed 14 May, 2024). Table~\ref{table:data_url} shows short URL addresses where readers can find the descriptions and trajectory files of the presented experiment setups. 

\begin{table}
	\normalsize                       %
	\setlength{\tabcolsep}{6pt}       % general space between columns (6pt standard) 
	\renewcommand{\arraystretch}{1.2} % general space between rows (1 standard)
	\centering
	\caption{Short URL addresses of the presented experiment setups (last accessed 14 May, 2024).}
	\label{table:data_url}
	\resizebox{7cm}{!}{
		\begin{tabular}{cl}
			\hline
			\hline
			Setup & Short URL\\
			\hline
			Uni-directional & http://ped.fz-juelich.de/da/2013unidirectional\\
			Bi-directional & http://ped.fz-juelich.de/da/2013bidirectional\\
			2-way crossing & http://ped.fz-juelich.de/da/2013crossing90\\
			3-way crossing & http://ped.fz-juelich.de/da/2013crossing120\\
			4-way crossing & http://ped.fz-juelich.de/da/2013crossing90\\
			\hline
			\hline
		\end{tabular}		
	}
	\vspace{-0.0cm}
\end{table}

The data processing, simulations, and analysis were carried out in Python with the help of open-source libraries. The code used for our study is publicly available at \url{https://doi.org/10.5281/zenodo.10455825}.

\section{Basic descriptive statistics of experiment scenarios}
\label{sec:appendix_basic_statistics}
In this section, we show basic descriptive statistics of experiment scenarios same as Table~\ref{table:basic_statistics} for different pedestrian flow setups, see Table~\ref{table:basic_statistics_all}. 

\begin{table*}
\normalsize                       %
\setlength{\tabcolsep}{6pt}       % general space between columns (6pt standard) 
\renewcommand{\arraystretch}{1.2} % general space between rows (1 standard)
\centering
\caption{Basic descriptive statistics of experiment scenarios.}
\label{table:basic_statistics_all}
\resizebox{14.0cm}{!}{
\begin{tabular}{c*{9}{r}}
\hline
\hline
& & & & & \multicolumn{4}{c}{No. contacts}\\
\cline{6-9}
Setup & fps & Scenario name & N & Period (s) & Parallel & Head-on & Crossing & total\\
\hline
Uni-directional & 25 & uni-01 & 148 & 75.52 & 644 (100\%) & 0 & 0 & 644\\
Uni-directional & 25 & uni-02 & 760 & 191.68 & 8220 (100\%) & 0 & 0 & 8220\\
Uni-directional & 25 & uni-03 & 916 & 161.84 & 17396 (100\%) & 0 & 0 & 17396\\
Uni-directional & 25 & uni-04 & 909 & 163.28 & 23214 (100\%) & 0 & 0 & 23214\\
Uni-directional & 25 & uni-05 & 905 & 157.68 & 25260 (100\%) & 0 & 0 & 25260\\
Uni-directional & 25 & uni-06 & 913 & 170.76 & 27948 (100\%) & 0 & 0 & 27948\\
Uni-directional & 25 & uni-07 & 914 & 205.08 & 34696 (100\%) & 0 & 0 & 34696\\
Uni-directional & 25 & uni-08 & 477 & 118.04 & 22164 (100\%) & 0 & 0 & 22164\\
Uni-directional & 25 & uni-09 & 310 & 78.92 & 16378 (100\%) & 0 & 0 & 16378\\
\hline
Bi-directional & 16 & bi-a1 & 377 & 95.38 & 5438 (26.86\%) & 14810 (73.14\%) & 0 & 20248\\
Bi-directional & 16 & bi-a2 & 522 & 143.94 & 13460 (30.20\%) & 31114 (69.80\%) & 0 & 44574\\
Bi-directional & 16 & bi-a3 & 542 & 143.00 & 15718 (30.22\%) & 36302 (69.78\%) & 0 & 52020\\
Bi-directional & 16 & bi-a4 & 544 & 165.31 & 18086 (30.09\%) & 42024 (69.91\%) & 0 & 60110\\
Bi-directional & 16 & bi-a5 & 560 & 177.31 & 17190 (26.63\%) & 47372 (73.37\%) & 0 & 64562\\
Bi-directional & 16 & bi-b01 & 141 & 118.88 & 386 (21.40\%) & 1418 (78.60\%) & 0 & 1804\\
Bi-directional & 16 & bi-b02 & 259 & 145.56 & 1148 (20.29\%) & 4510 (79.71\%) & 0 & 5658\\
Bi-directional & 16 & bi-b03 & 480 & 202.88 & 4088 (24.33\%) & 12712 (75.67\%) & 0 & 16800\\
Bi-directional & 16 & bi-b04 & 743 & 296.50 & 15586 (25.04\%) & 46654 (74.96\%) & 0 & 62240\\
Bi-directional & 16 & bi-b05 & 643 & 269.75 & 14192 (22.43\%) & 49080 (77.57\%) & 0 & 63272\\
Bi-directional & 16 & bi-b06 & 830 & 388.81 & 24804 (26.02\%) & 70524 (73.98\%) & 0 & 95328\\
Bi-directional & 16 & bi-b07 & 606 & 254.56 & 19208 (25.71\%) & 55508 (74.29\%) & 0 & 74716\\
Bi-directional & 16 & bi-b08 & 703 & 359.69 & 23202 (24.44\%) & 71738 (75.56\%) & 0 & 94940\\
Bi-directional & 16 & bi-b09 & 483 & 186.38 & 6426 (28.54\%) & 16088 (71.46\%) & 0 & 22514\\
Bi-directional & 16 & bi-b10 & 736 & 324.56 & 17154 (29.81\%) & 40398 (70.19\%) & 0 & 57552\\
\hline
2-way crossing & 16 & crossing-90-d01 & 603 & 199.06 & 5248 (27.74\%) & 0 & 13672 (72.26\%) & 18920\\
2-way crossing & 16 & crossing-90-d02 & 604 & 192.00 & 6722 (46.34\%) & 0 & 7784 (53.66\%) & 14506\\
2-way crossing & 16 & crossing-90-d03 & 606 & 186.69 & 2674 (24.79\%) & 0 & 8114 (75.21\%) & 10788\\
2-way crossing & 16 & crossing-90-d05 & 600 & 153.75 & 5404 (26.06\%) & 0 & 15330 (73.94\%) & 20734\\
2-way crossing & 16 & crossing-90-d06 & 597 & 131.56 & 15582 (51.03\%) & 0 & 14952 (48.97\%) & 30534\\
2-way crossing & 16 & crossing-90-d07 & 604 & 139.50 & 10568 (28.48\%) & 0 & 26544 (71.52\%) & 37112\\
2-way crossing & 16 & crossing-90-d08 & 592 & 147.88 & 22418 (53.39\%) & 0 & 19574 (46.61\%) & 41992\\
\hline
3-way crossing & 16 & crossing-120-a1 & 254 & 78.88 & 1156 (21.85\%) & 4134 (78.15\%) & 0 & 5290\\
3-way crossing & 16 & crossing-120-a2 & 286 & 93.56 & 1428 (27.20\%) & 3822 (72.80\%) & 0 & 5250\\
3-way crossing & 16 & crossing-120-a3 & 341 & 96.56 & 3638 (26.57\%) & 10054 (73.43\%) & 0 & 13692\\
3-way crossing & 16 & crossing-120-a4 & 710 & 148.69 & 14714 (26.64\%) & 40514 (73.36\%) & 0 & 55228\\
3-way crossing & 16 & crossing-120-a5 & 814 & 186.12 & 27026 (27.41\%) & 71578 (72.59\%) & 0 & 98604\\
3-way crossing & 16 & crossing-120-a6 & 783 & 182.44 & 29764 (30.62\%) & 67456 (69.38\%) & 0 & 97220\\
3-way crossing & 16 & crossing-120-a7 & 886 & 205.06 & 33904 (28.90\%) & 83406 (71.10\%) & 0 & 117310\\
3-way crossing & 16 & crossing-120-b1 & 769 & 215.63 & 34654 (36.80\%) & 59518 (63.20\%) & 0 & 94172\\
3-way crossing & 16 & crossing-120-b2 & 700 & 182.38 & 29636 (37.93\%) & 48492 (62.07\%) & 0 & 78128\\
3-way crossing & 16 & crossing-120-c1 & 262 & 86.81 & 1758 (36.76\%) & 3024 (63.24\%) & 0 & 4782\\
3-way crossing & 16 & crossing-120-c2 & 382 & 99.12 & 7184 (36.38\%) & 12564 (63.62\%) & 0 & 19748\\
3-way crossing & 16 & crossing-120-c3 & 573 & 156.75 & 18016 (39.00\%) & 28176 (61.00\%) & 0 & 46192\\
3-way crossing & 16 & crossing-120-c4 & 697 & 196.75 & 28546 (39.33\%) & 44040 (60.67\%) & 0 & 72586\\
3-way crossing & 16 & crossing-120-c5 & 627 & 201.81 & 30788 (40.41\%) & 45402 (59.59\%) & 0 & 76190\\
3-way crossing & 16 & crossing-120-c6 & 811 & 274.50 & 38872 (39.89\%) & 58574 (60.11\%) & 0 & 97446\\
3-way crossing & 16 & crossing-120-c7 & 842 & 191.50 & 54378 (42.60\%) & 73270 (57.40\%) & 0 & 127648\\
\hline
4-way crossing & 25 & crossing-90-a01 & 247 & 168.08 & 68 (3.85\%) & 920 (52.04\%) & 780 (44.12\%) & 1768\\
4-way crossing & 25 & crossing-90-a02 & 439 & 132.28 & 3186 (15.66\%) & 7566 (37.19\%) & 9594 (47.15\%) & 20346\\
4-way crossing & 25 & crossing-90-a03 & 323 & 90.76 & 2816 (15.18\%) & 6718 (36.22\%) & 9014 (48.60\%) & 18548\\
4-way crossing & 25 & crossing-90-a04 & 352 & 105.04 & 4636 (13.61\%) & 11712 (34.38\%) & 17716 (52.01\%) & 34064\\
4-way crossing & 25 & crossing-90-a05 & 337 & 88.08 & 6210 (16.83\%) & 11912 (32.29\%) & 18770 (50.88\%) & 36892\\
4-way crossing & 25 & crossing-90-a06 & 269 & 47.56 & 3234 (13.12\%) & 6150 (24.96\%) & 15260 (61.92\%) & 24644\\
4-way crossing & 25 & crossing-90-a07 & 323 & 100.04 & 2338 (15.60\%) & 5564 (37.13\%) & 7082 (47.26\%) & 14984\\
4-way crossing & 25 & crossing-90-a08 & 298 & 71.96 & 7850 (25.11\%) & 7482 (23.93\%) & 15928 (50.95\%) & 31260\\
4-way crossing & 25 & crossing-90-a09 & 299 & 116.24 & 310 (7.03\%) & 2066 (46.83\%) & 2036 (46.15\%) & 4412\\
4-way crossing & 25 & crossing-90-a10 & 324 & 94.08 & 882 (9.60\%) & 3834 (41.73\%) & 4472 (48.67\%) & 9188\\
\hline
\hline
\end{tabular}		
}
\vspace{-0.0cm}
\end{table*}

\section{Summary of motion type classification results}
\label{sec:appendix_classification_results}
\begin{figure*}
	\centering
	\includegraphics[width=15.5cm]{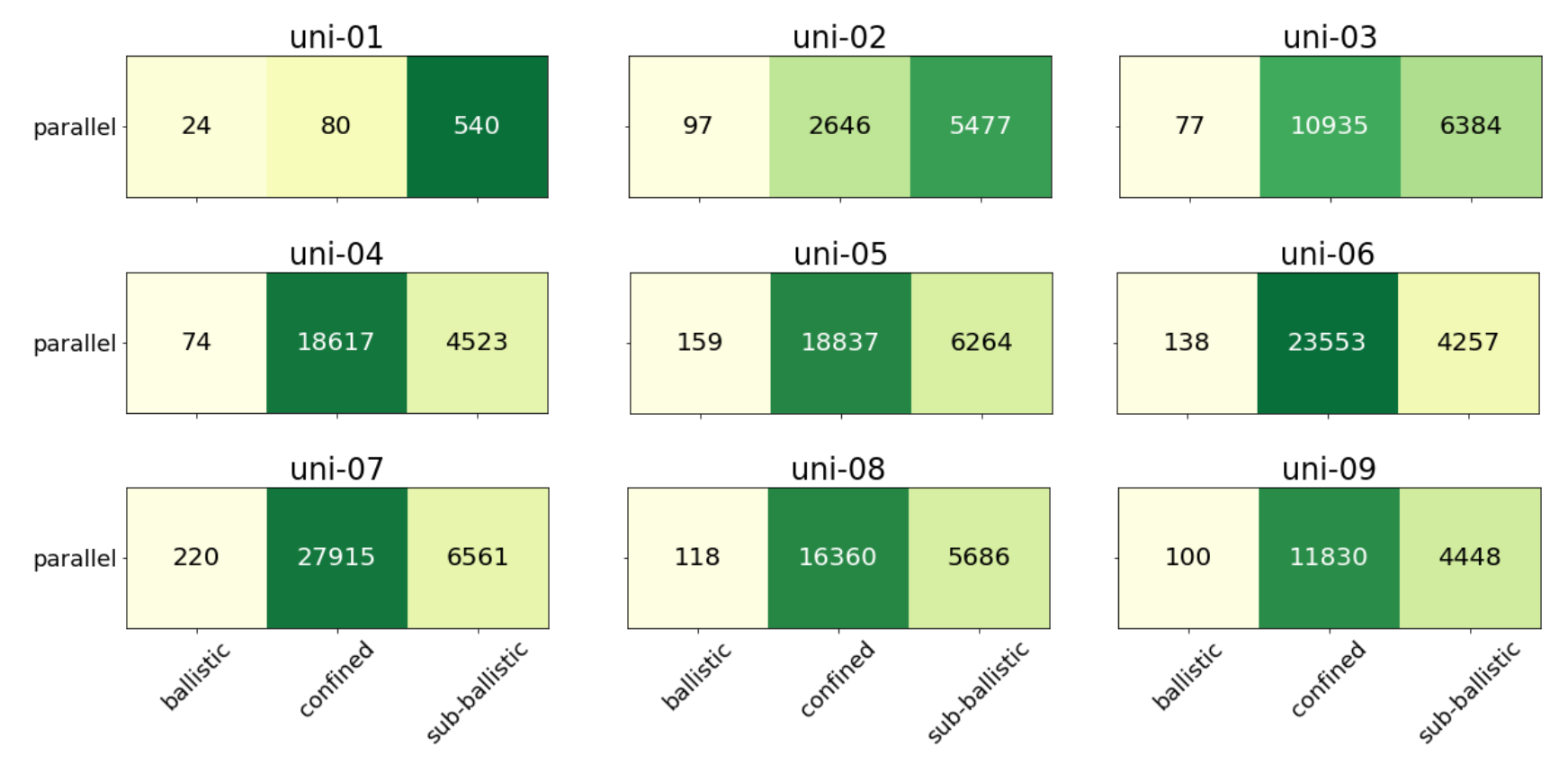}
	\caption{Summary of motion type classification results for unidirectional flow scenarios.} 
	\label{fig:motion_type_classification_uni}
\end{figure*}

\begin{figure*}
	\centering
	\includegraphics[width=15.5cm]{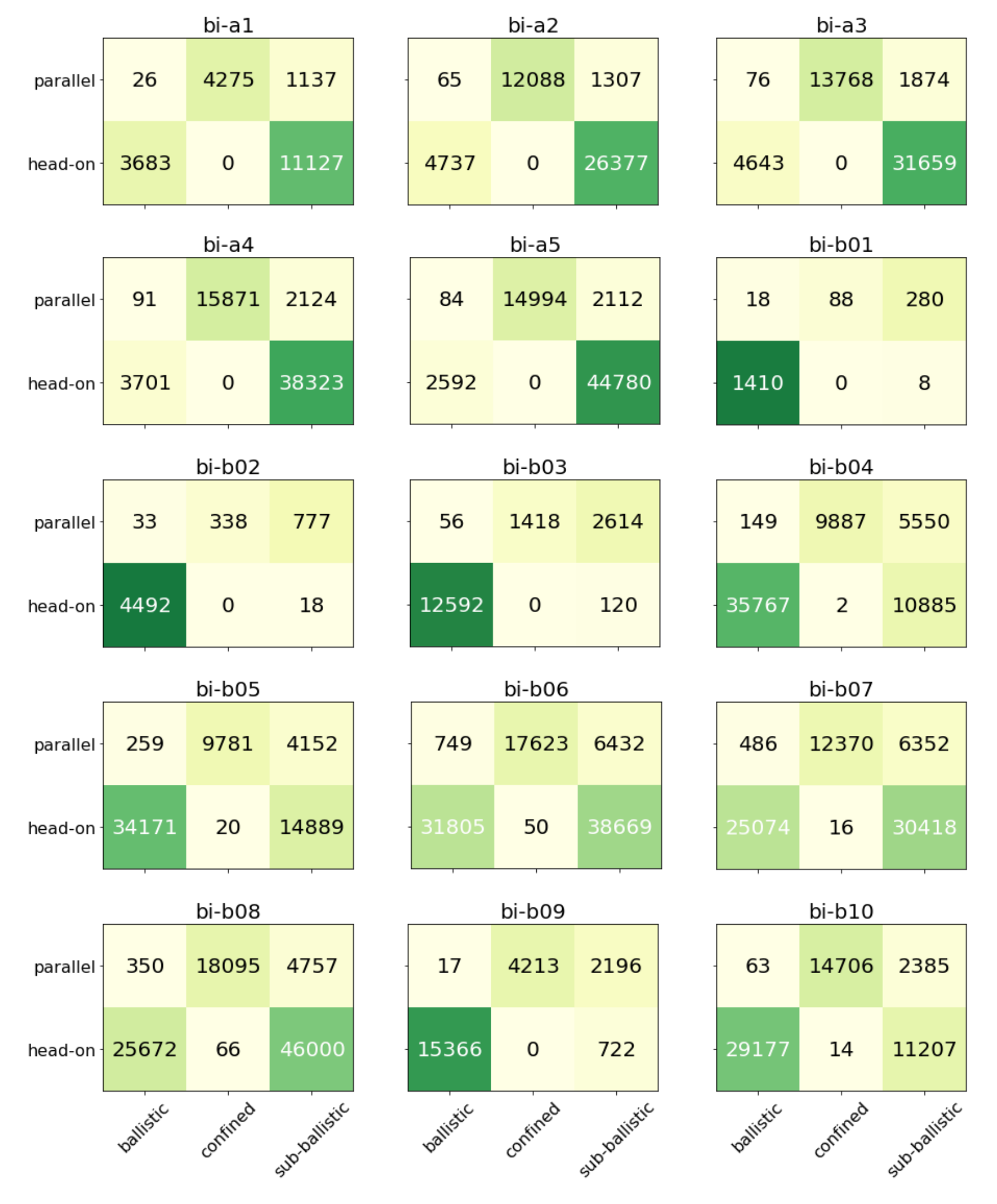}
	\caption{Summary of motion type classification results for bidirectional flow scenarios.} 
	\label{fig:motion_type_classification_bi}
\end{figure*}

\begin{figure*}
	\centering
	\includegraphics[width=15.5cm]{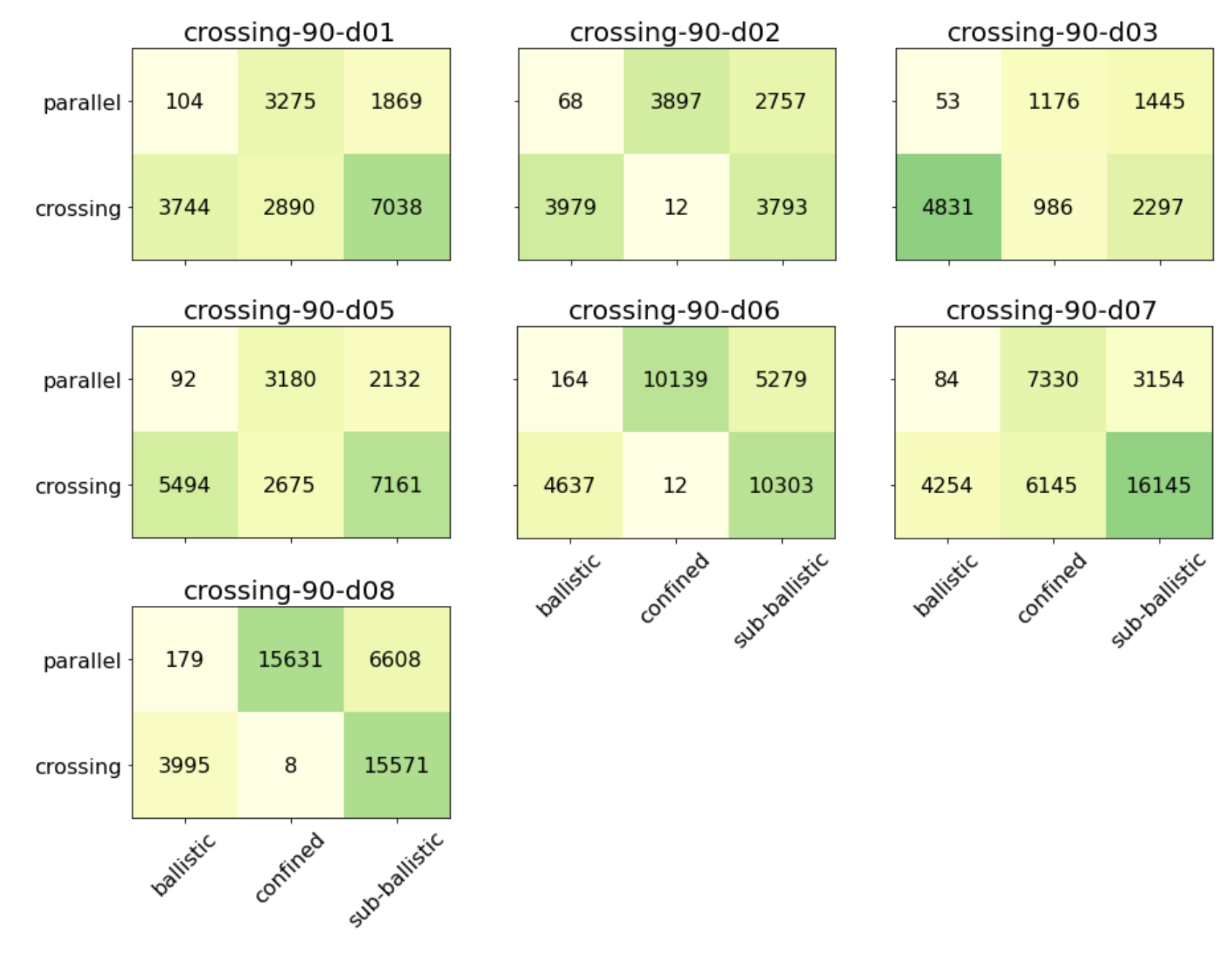}
	\caption{Summary of motion type classification results for 2-way crossing flow scenarios.} 
	\label{fig:motion_type_classification_2way}
\end{figure*}

\begin{figure*}
	\centering
	\includegraphics[width=15.5cm]{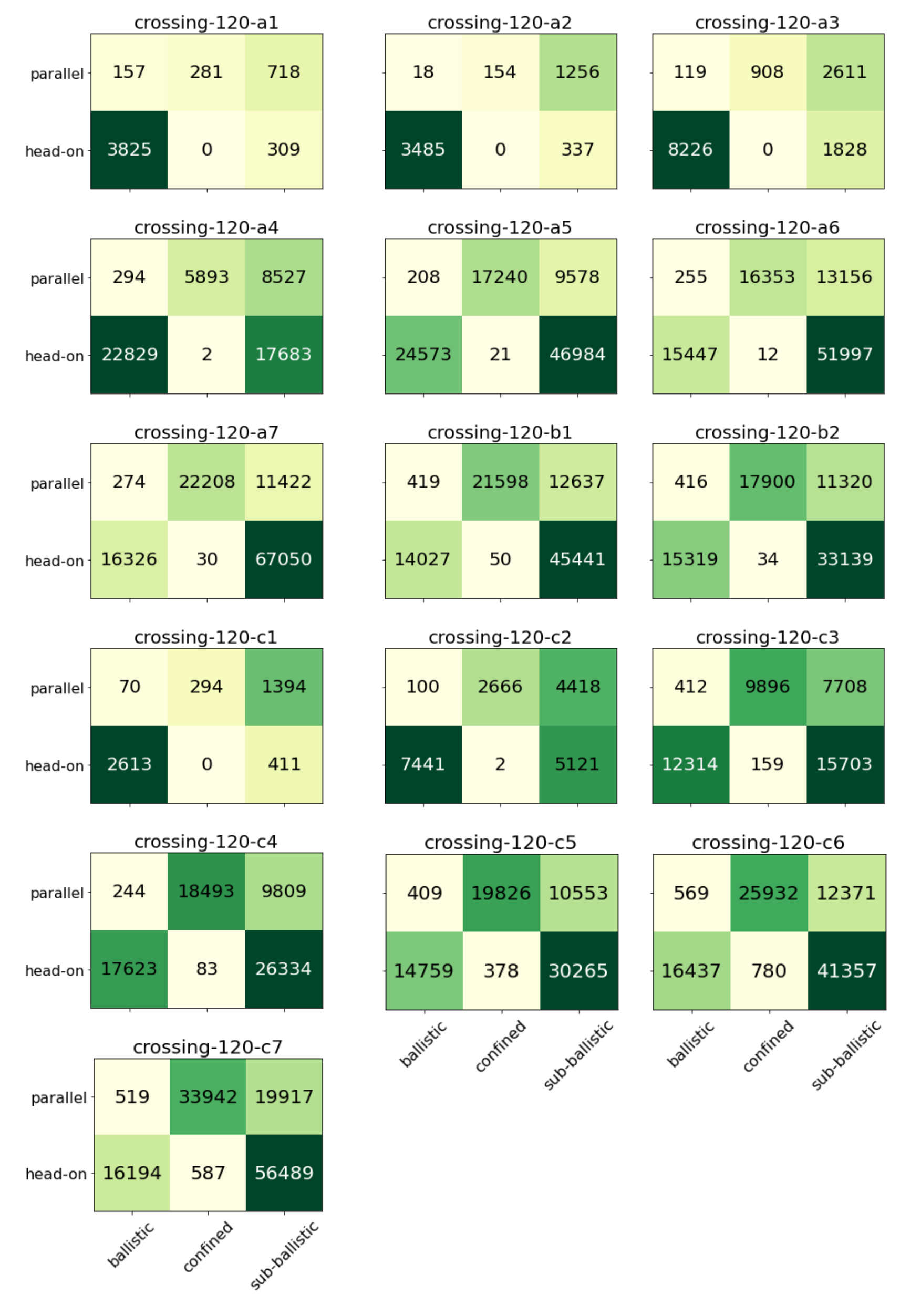}
	\caption{Summary of motion type classification results for 3-way crossing flow scenarios.} 
	\label{fig:motion_type_classification_3way}
\end{figure*}

\begin{figure*}
	\centering
	\includegraphics[width=15.5cm]{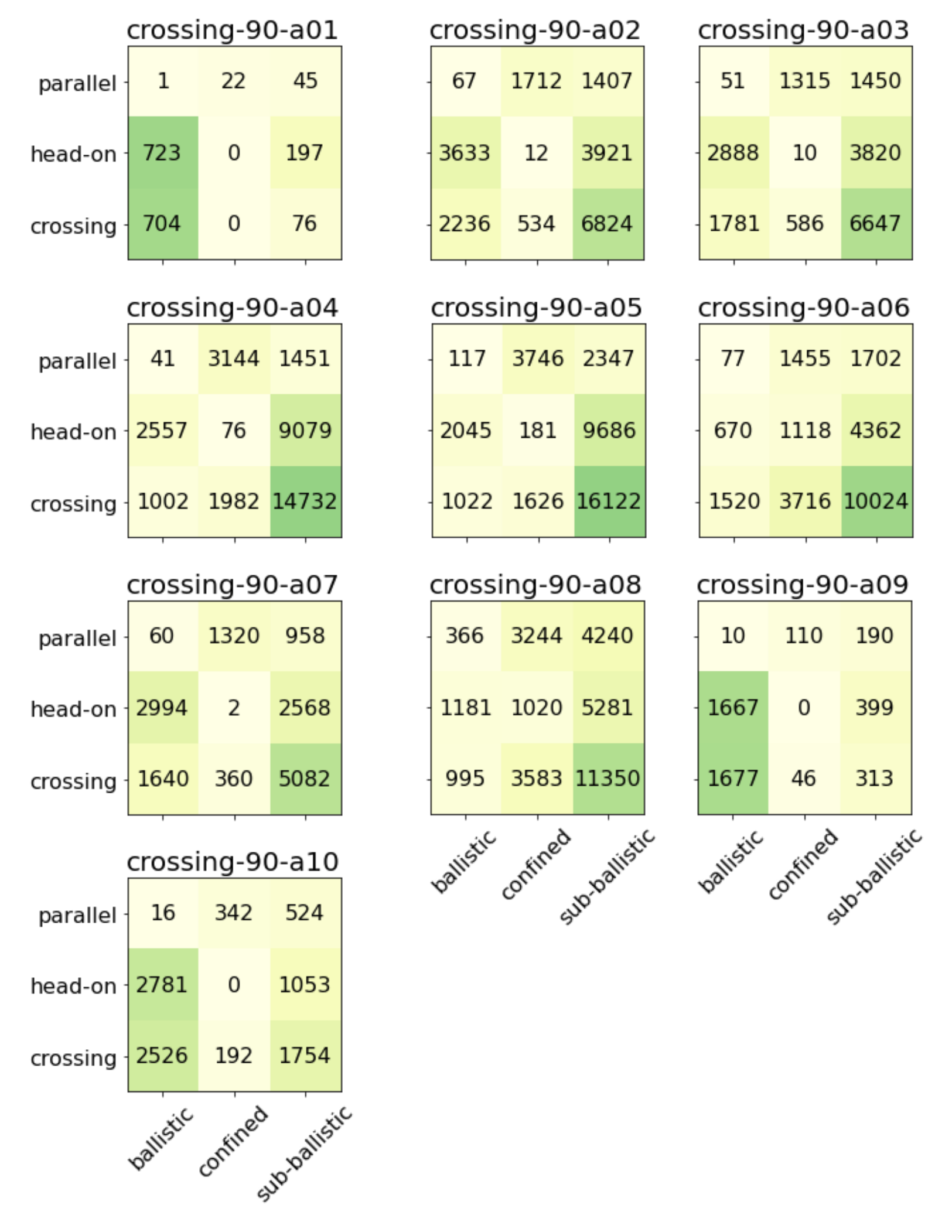}
	\caption{Summary of motion type classification results for 4-way crossing flow scenarios.} 
	\label{fig:motion_type_classification_4way}
\end{figure*}

In this section, we present summary of motion type classification results same as Fig.~\ref{fig:motion_type_classification} for different pedestrian flow setups: uni-directional (Fig.~\ref{fig:motion_type_classification_uni}), bi-directional (Fig.~\ref{fig:motion_type_classification_bi}), 2-way crossing (Fig.~\ref{fig:motion_type_classification_2way}), 3-way crossing (Fig.~\ref{fig:motion_type_classification_3way}), and 4-way crossing flows (Fig.~\ref{fig:motion_type_classification_4way}). 

\section{Average contact duration}
\label{sec:appendix_tc_avg}
\begin{figure*}
	\centering
	\includegraphics[width=15.5cm]{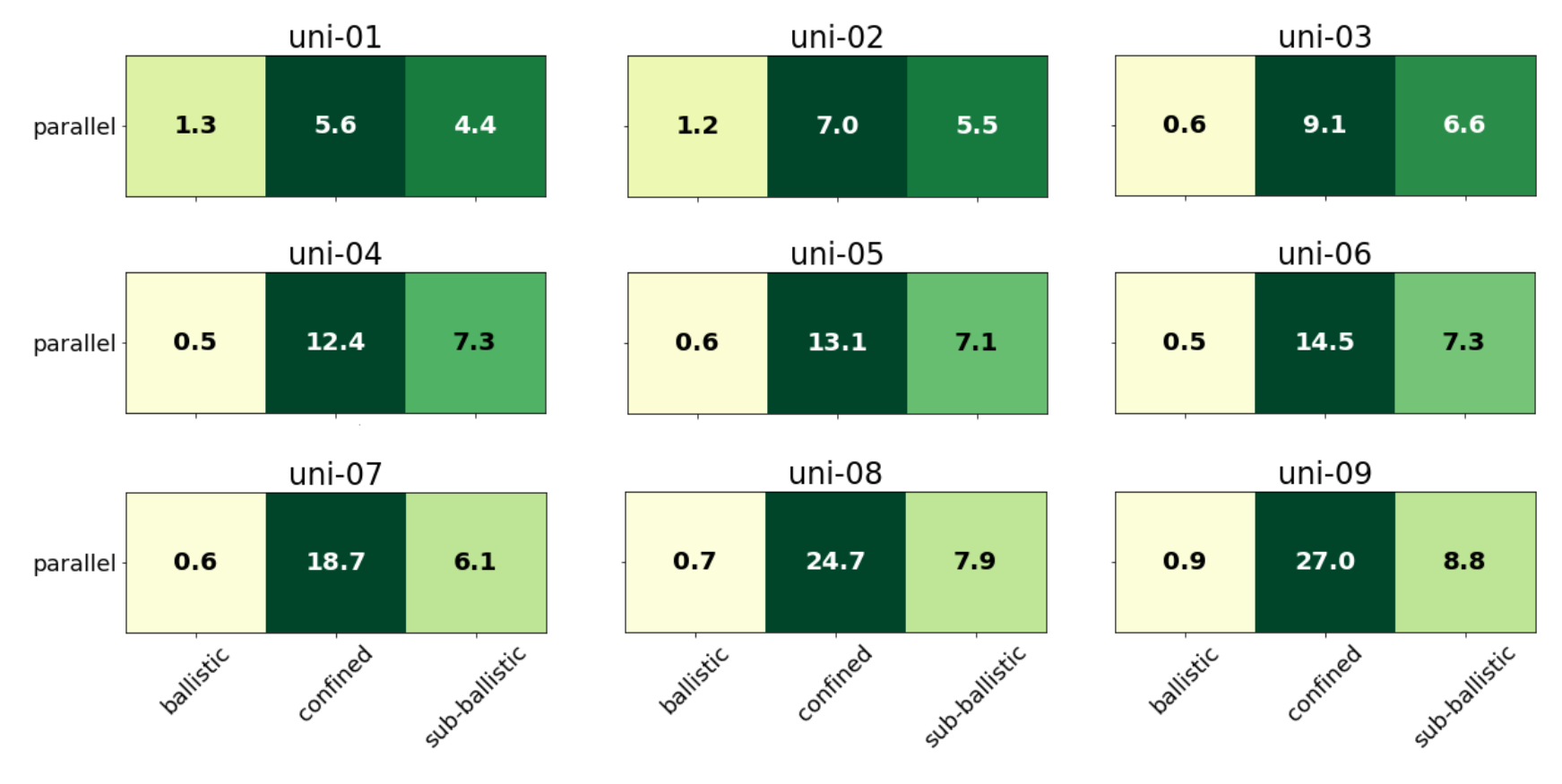}
	\caption{Average values of contact duration $t_c$ for unidirectional flow scenarios.} 
	\label{fig:motion_type_tc_uni}
\end{figure*}

\begin{figure*}
	\centering
	\includegraphics[width=15.5cm]{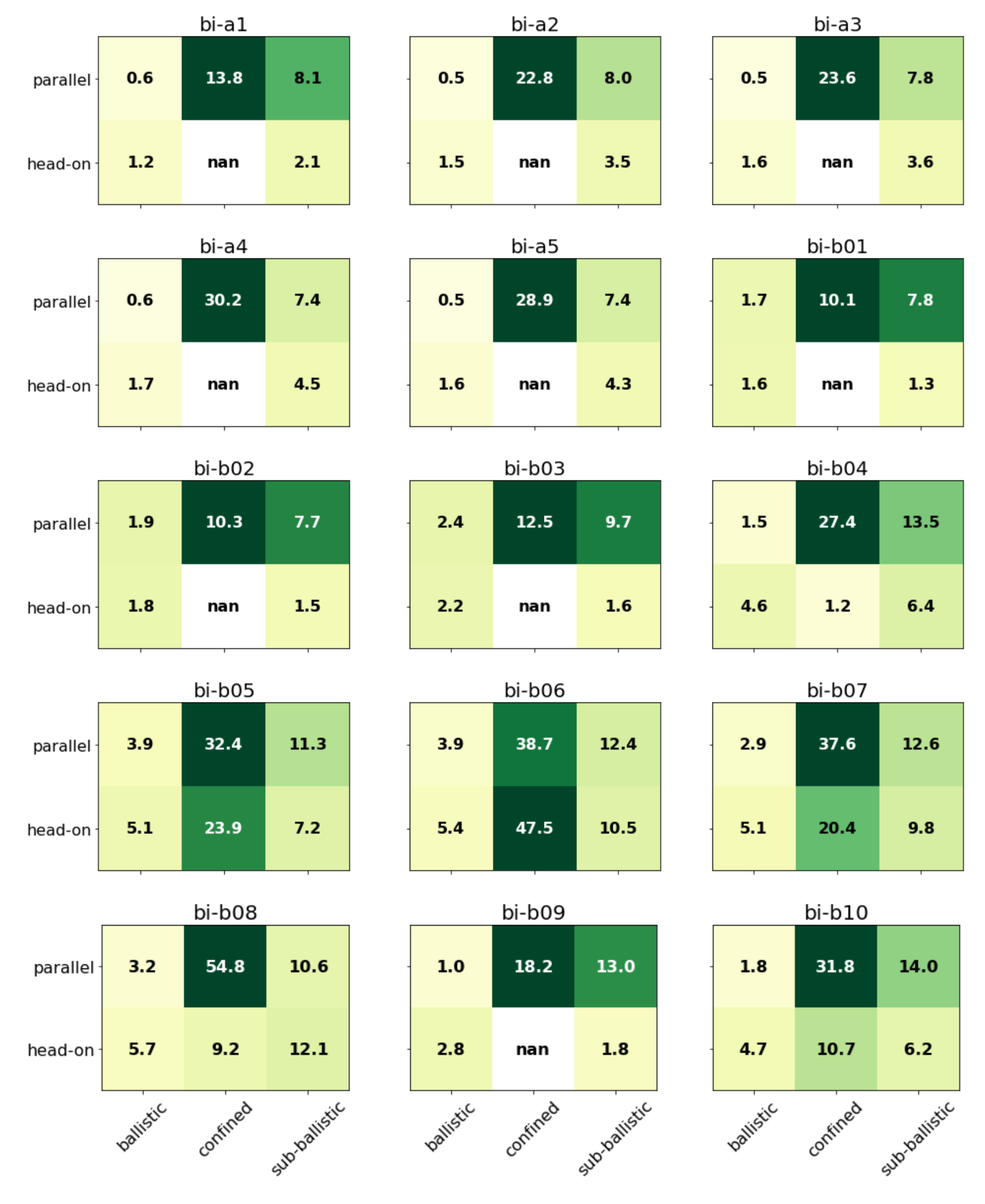}
	\caption{Average values of contact duration $t_c$ for bidirectional flow scenarios.} 
	\label{fig:motion_type_tc_bi}
\end{figure*}

\begin{figure*}
	\centering
	\includegraphics[width=15.5cm]{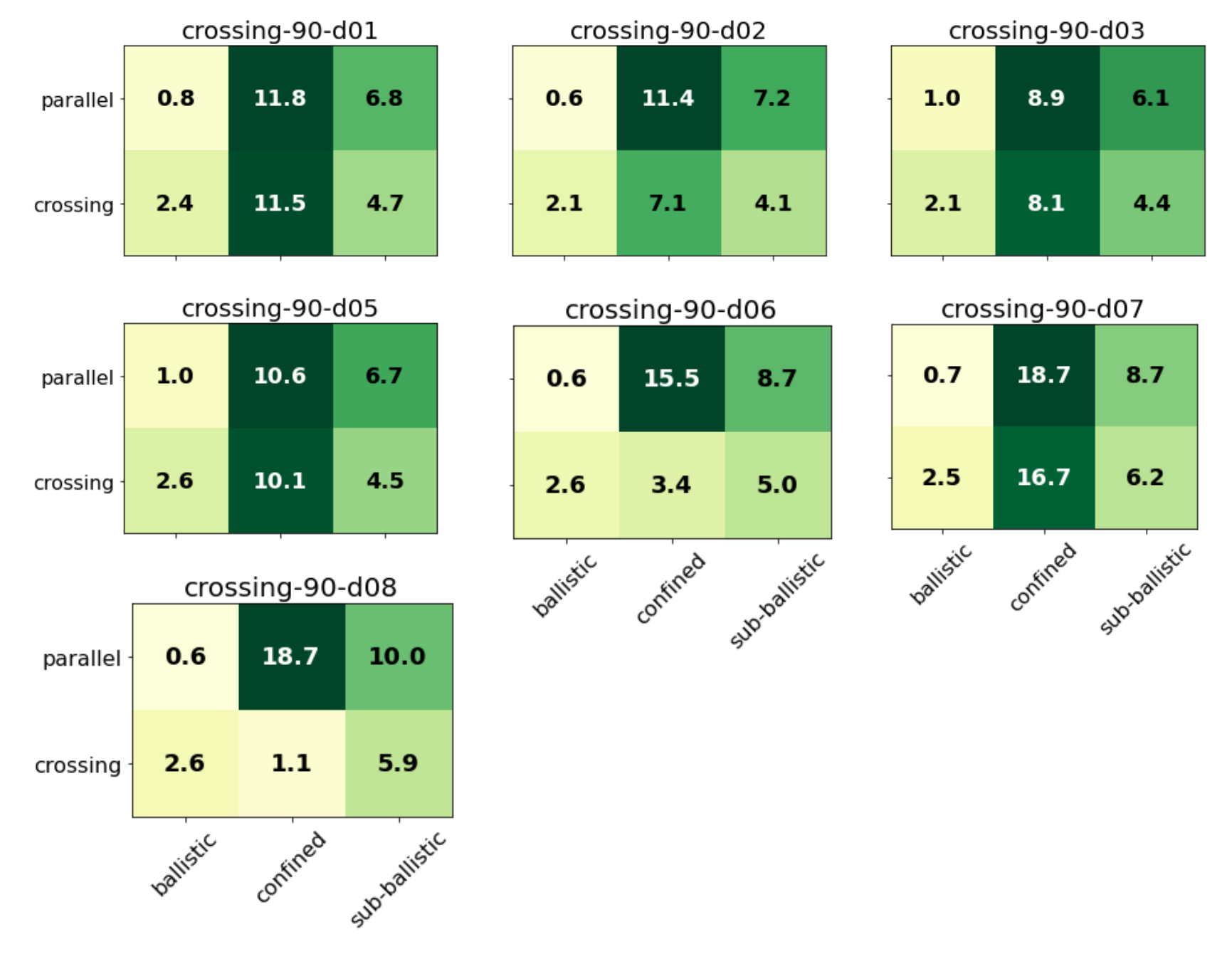}
	\caption{Average values of contact duration $t_c$ for 2-way crossing flow scenarios.} 
	\label{fig:motion_type_tc_2way}
\end{figure*}

\begin{figure*}
	\centering
	\includegraphics[width=15.5cm]{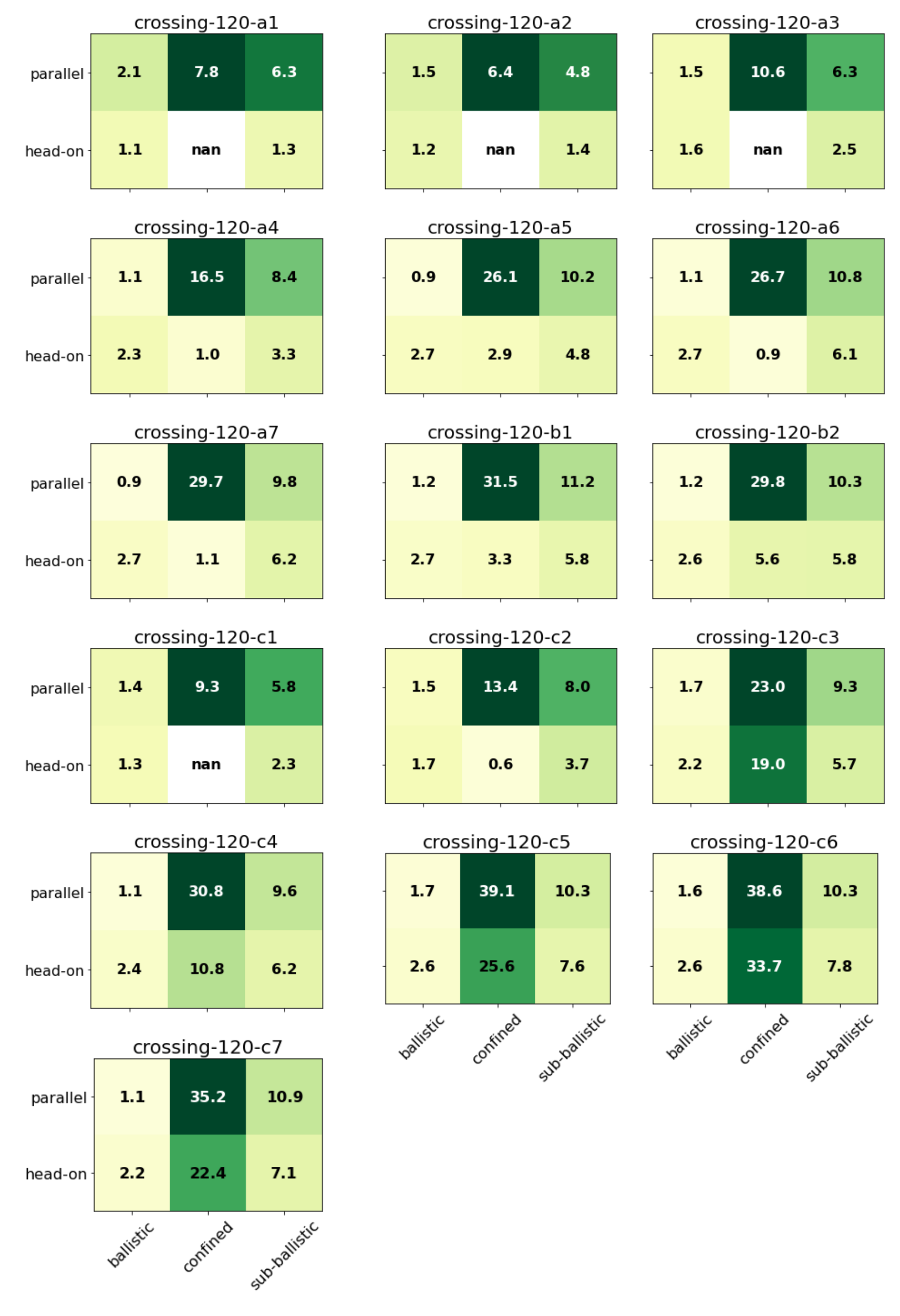}
	\caption{Average values of contact duration $t_c$ for 3-way crossing flow scenarios.} 
	\label{fig:motion_type_tc_3way}
\end{figure*}

\begin{figure*}
	\centering
	\includegraphics[width=15.5cm]{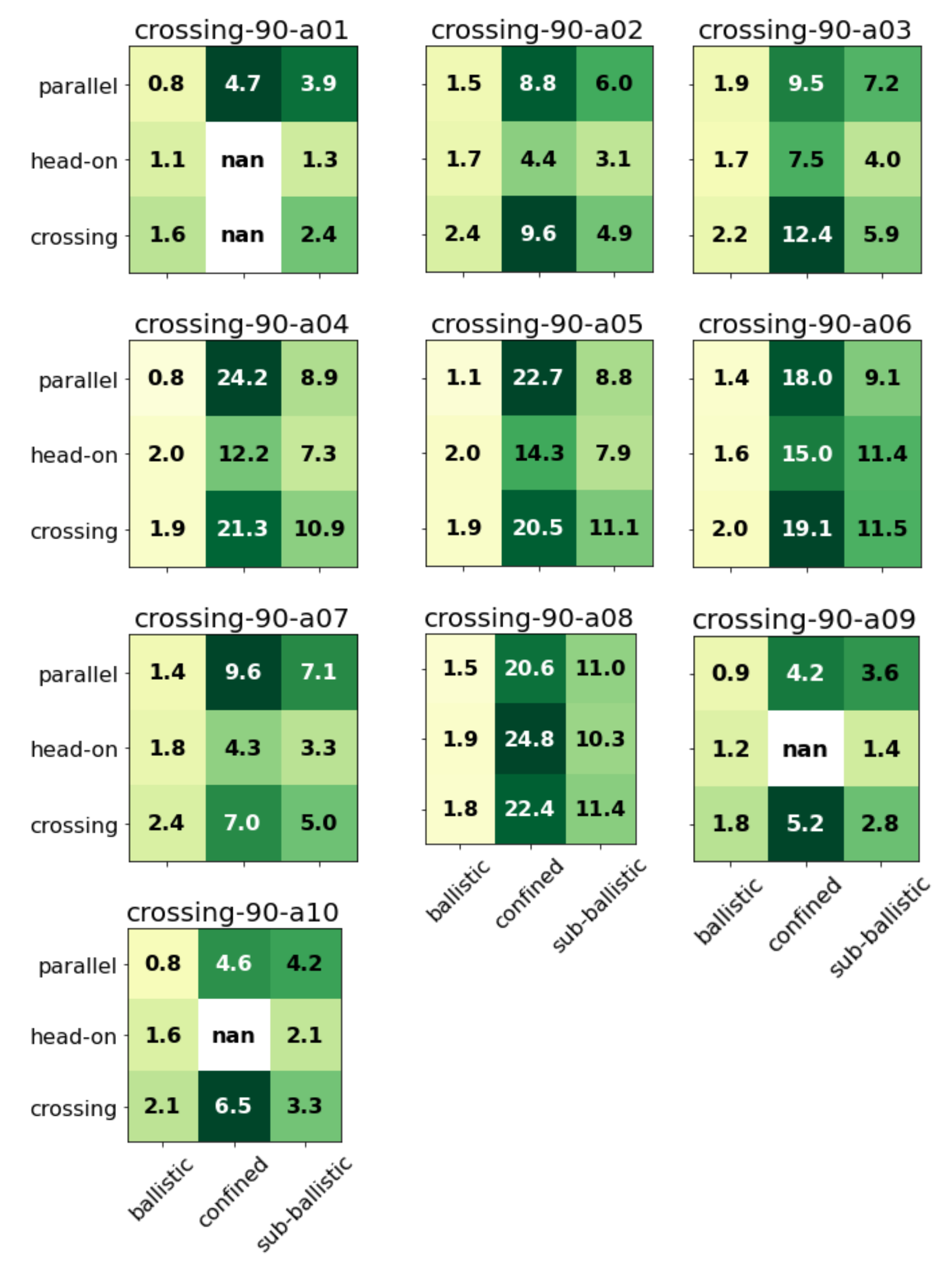}
	\caption{Average values of contact duration $t_c$ for 4-way crossing flow scenarios.} 
	\label{fig:motion_type_tc_4way}
\end{figure*}

In this section, we illustrate average values of contact duration $t_c$ same as Fig.~\ref{fig:motion_type_tc} for different pedestrian flow setups: uni-directional (Fig.~\ref{fig:motion_type_tc_uni}), bi-directional (Fig.~\ref{fig:motion_type_tc_bi}), 2-way crossing (Fig.~\ref{fig:motion_type_tc_2way}), 3-way crossing (Fig.~\ref{fig:motion_type_tc_3way}), and 4-way crossing flows (Fig.~\ref{fig:motion_type_tc_4way}). 

% \newpage
\clearpage\clearpage
\bibliographystyle{model1-num-names}
%\bibliographystyle{cas-model2-names}
%%\bibliography{reference_20200210a7}

\begin{thebibliography}{}
% ballistic movement in ad-hoc networks
\bibitem{Samar_IEEE2006}	
Samar, P.M, Wicker, S.B.: Link dynamics and protocol design in a multihop mobile environment. IEEE Transactions on Mobile Computing \textbf{5}, pp. 1156--1172 (2006).

% ballistic movement in ad-hoc networks
\bibitem{Wu_IEEE2008}	
Wu, Y.T., Liao, W., Tsao, C.L., Lin, T.N.: Impact of node mobility on link duration in multihop mobile networks. IEEE Transactions on Vehicular Technology \textbf{58}, pp. 2435--2442 (2008).

% contact interaction modeling and applications
\bibitem{Hu_2013}	
Hu, H., Nigmatulina, K., Eckhoff, P.: The scaling of contact rates with population density for the infectious disease models. Mathematical Biosciences \textbf{244}, pp. 125--134 (2013).

% contact interaction modeling and applications
\bibitem{Manlove_2022}	
Manlove, K., Wilber, M., White, L., Bastille‐Rousseau, G., Yang, A., Gilbertson, M.L., Craft, M.E., Cross, P.C., Wittemyer, G., Pepin, K.M.: Defining an epidemiological landscape that connects movement ecology to pathogen transmission and pace‐of‐life. Ecology Letters \textbf{25}, pp. 1760-1782 (2022).

% ballistic movement
\bibitem{Rast_PRE2022}	
Rast, M.P.: Contact statistics in populations of noninteracting random walkers in two dimensions. Physical Review E \textbf{105}, 014103 (2022).

% relevant literature on single particle tracking and trajectory analysis
\bibitem{Saxton_1997}	
Saxton, M.J., Jacobson, K.: Single-particle tracking: applications to membrane dynamics. Annual Review of Biophysics and Biomolecular Structure  \textbf{26}, pp. 373--399 (1997).

% relevant literature on single particle tracking and trajectory analysis
\bibitem{Manzo_2015}	
Manzo, C., Garcia-Parajo, M.F.: A review of progress in single particle tracking: from methods to biophysical insights. Reports on Progress in Physics \textbf{78}, 124601 (2015).

% relevant literature on single particle tracking and trajectory analysis
\bibitem{Shen_2017}	
Shen, H., Tauzin, L.J., Baiyasi, R., Wang, W., Moringo, N., Shuang, B., Landes, C.F.: Single particle tracking: from theory to biophysical applications. Chemical Reviews \textbf{117}, pp. 7331--7376 (2017).

% relevant literature in movement ecology
\bibitem{Benhamou_2007}	
Benhamou, S.: How many animals really do the L\'{e}vy walk?. Ecology \textbf{88}, pp. 1962--1969 (2007).

% relevant literature in movement ecology
\bibitem{Edelhoff_2016}	
Edelhoff, H., Signer, J., Balkenhol, N.: Path segmentation for beginners: an overview of current methods for detecting changes in animal movement patterns. Movement Ecology \textbf{4}, 21 (2016).

% relevant literature in movement ecology
\bibitem{Getz_PNAS2008}	
Getz, W.M., Saltz, D.: A framework for generating and analyzing movement paths on ecological landscapes. Proceedings of the National Academy of Sciences \textbf{105}, pp. 19066--19071 (2008).

% contact radius
\bibitem{Rutten_SciRep2022}
Rutten, P., Lees, M.H., Klous, S., Heesterbeek, H., Sloot, P.: Modelling the dynamic relationship between spread of infection and observed crowd movement patterns at large scale events. Scientific Reports \textbf{12}, 14825 (2022).

% relevant literature in epidemiology 
\bibitem{Wilber_2022}	
Wilber, M.Q., Yang, A., Boughton, R., Manlove, K.R., Miller, R.S., Pepin, K.M., Wittemyer, G.: A model for leveraging animal movement to understand spatio‐temporal disease dynamics. Ecology Letters \textbf{25}, pp. 1290--1304 (2022).

% relevant literature in single particle tracking (SPT) analysis
\bibitem{Qian_1991}	
Qian, H., Sheetz, M.P., Elson, E.L.: Single particle tracking. Analysis of diffusion and flow in two-dimensional systems. Biophysical Journal \textbf{60}, pp. 910--921 (1991).

% relevant literature in mean square distance (MSD)
\bibitem{Michalet_PRE2010}	
Michalet, X.: Mean square displacement analysis of single-particle trajectories with localization error: Brownian motion in an isotropic medium. Physical Review E \textbf{82}, 041914 (2010).

% relevant literature in mean square distance (MSD)
\bibitem{Goulian_2000}	
Goulian, M., Simon, S.M.: Tracking single proteins within cells. Biophysical Journal \textbf{79}, pp. 2188--2198.

% relevant literature in mean square distance (MSD)
\bibitem{Hubicka_PRE2020}	
Hubicka, K., Janczura, J.: Time-dependent classification of protein diffusion types: A statistical detection of mean-squared-displacement exponent transitions. Physical Review E \textbf{1010}, 022107.

% diffusion of individual movements perpendicular to the flow direction, application of MSD
\bibitem{Murakami_interface2019}
Murakami, H., Feliciani, C., Nishinari, K.: L\'{e}vy walk process in self-organization of pedestrian crowds. Journal of the Royal Society Interface \textbf{16}, 20180939 (2019).

\bibitem{Murakami_SciAdv2021}
Murakami, H., Feliciani, C., Nishiyama, Y., Nishinari, K.: Mutual anticipation can contribute to self-organization in human crowds. Science Advances \textbf{7}, eabe7758 (2021).

% T_n, limitations of MSD
\bibitem{Briane_2016}
Briane V, Vimond M, Kervrann C.: An adaptive statistical test to detect non Brownian diffusion from particle trajectories. In: 2016 IEEE 13th International Symposium on Biomedical Imaging (ISBI), pp. 972--975, IEEE, Prague, Czech Republic (2016).

% T_n, limitations of MSD
\bibitem{Briane_PRE2018}
Briane, V., Kervrann, C., Vimond, M.: Statistical analysis of particle trajectories in living cells. Physical Review E \textbf{97}, 062121 (2018).

% feature extraction
\bibitem{Janczura_PRE2020}
Janczura, J., Kowalek, P., Loch-Olszewska, H., Szwabi\'{n}ski, J., Weron, A.: Classification of particle trajectories in living cells: Machine learning versus statistical testing hypothesis for fractional anomalous diffusion. Physical Review E \textbf{102}, 032402 (2020).

% relevant literature in mean square distance (MSD)
\bibitem{Kepten_PLOS2015}	
Kepten, E., Weron, A., Sikora, G., Burnecki, K., Garini, Y.: Guidelines for the fitting of anomalous diffusion mean square displacement graphs from single particle tracking experiments. PLOS One \textbf{10}, e0117722 (2015).

% relevant literature in mean square distance (MSD)
\bibitem{Burnecki_SciRep2015}	
Burnecki, K., Kepten, E., Garini, Y., Sikora, G., Weron, A.: Estimating the anomalous diffusion exponent for single particle tracking data with measurement errors-An alternative approach. Scientific Reports \textbf{10}, 11306 (2015).

% statistical testing approach
\bibitem{Weron_PRE2019}
Weron, A., Janczura, J., Boryczka, E., Sungkaworn, T., Calebiro, D.: Statistical testing approach for fractional anomalous diffusion classification. Physical Review E \textbf{99}, 042149 (2019).

% relevant literature in mean square distance (MSD)
\bibitem{Janczura_2022}	
Janczura, J., Burnecki, K., Muszkieta, M., Stanislavsky, A., Weron, A.: Classification of random trajectories based on the fractional L\'{e}vy stable motion. Chaos, Solitons \& Fractals \textbf{154}, 111606 (2022).

% feature extraction
\bibitem{Kowalek_JPhysA2022}
Kowalek, P., Loch-Olszewska, H., Łaszczuk, \L., Opała, J., Szwabiński, J.: Boosting the performance of anomalous diffusion classifiers with the proper choice of features. Journal of Physics A: Mathematical and Theoretical \textbf{55}, 244005 (2022).

% iccs paper
\bibitem{Kwak_iccs2023}
Kwak, J., Lees, M.H., Cai, W.: Characterization of pedestrian contact interaction trajectories. Lecture Notes in Computer Science \textbf{14073}, pp. 18--32 (2023).
%% ed Mikyška, J., de Mulatier, C., Paszynski, M., Krzhizhanovskaya, V.V., Dongarra, J.J., Sloot, P.M.

% experiment dataset
\bibitem{Holl_Dissertation2016}
Holl, S.: Methoden f\"{u}r die Bemessung der Leistungsf\"{a}higkeit multidirektional genutzter Fu{\ss}verkehrsanlagen. Bergische Universität, Wuppertal, Germany (2016).

% experiment dataset
\bibitem{Cao_JSTAT2017}
Cao, S., Seyfried, A., Zhang, J., Holl, S., Song, W.: Fundamental diagrams for multidirectional pedestrian flows. Journal of Statistical Mechanics: Theory and Experiment \textbf{2017}, 033404 (2017).

% experiment dataset
\bibitem{url_dataset}
Data archive of experimental data from studies about pedestrian dynamics. \url{https://ped.fz-juelich.de/da/doku.php?id=start#data_section}. Last accessed 14 May, 2024.

% contact radius
\bibitem{Han_Lancet2020}
Han, E., Tan, M.M.J., Turk, E., Sridhar, D., Leung, G.M., Shibuya, K., Asgari, N., Oh, J., García-Basteiro, A.L., Hanefeld, J., Cook, A.R.: Lessons learnt from easing COVID-19 restrictions: an analysis of countries and regions in Asia Pacific and Europe. The Lancet \textbf{396}, pp. 1525--1534 (2020).

% contact radius
\bibitem{Ronchi_SafetySci2020}
Ronchi, E., Lovreglio, R.: EXPOSED: An occupant exposure model for confined spaces to retrofit crowd models during a pandemic. Safety Science \textbf{130}, 104834 (2020).

% contact radius
\bibitem{Garcia_SafetySci2021}
Garcia, W., Mendez, S., Fray, B., Nicolas, A.: Model-based assessment of the risks of viral transmission in non-confined crowds. Safety Science \textbf{144}, 105453 (2021).

% contact radius
\bibitem{Mendez_AdvSci2023}
Mendez, S., Garcia, W. and Nicolas, A., 2023. From microscopic droplets to macroscopic crowds: Crossing the scales in models of short‐range respiratory disease transmission, with application to COVID‐19. Advanced Science \textbf{10}, 2205255 (2023).

% contact radius
\bibitem{Nicolas_2023}
Nicolas, A., Mendez, S.: Viral transmission in pedestrian crowds: Coupling an open-source code assessing the risks of airborne contagion with diverse pedestrian dynamics models. arXiv preprint arXiv:2312.01779 (2023). \url{https://arxiv.org/abs/2312.01779}.

% contact radius
\bibitem{Rahn_plos2022}
Rahn, S., G\"{o}del, M., K\"{o}ster, G. and Hofinger, G.: Modelling airborne transmission of SARS-CoV-2 at a local scale. PLOS One \textbf{17}, e0273820 (2022).

% 
\bibitem{Bale_SciRep2022}
Bale, R., Iida, A., Yamakawa, M., Li, C. and Tsubokura, M.: Quantifying the COVID-19 infection risk due to droplet/aerosol inhalation. Scientific Reports \textbf{12}, 11186 (2022).

% confined diffusion
\bibitem{Arinstein_PRE2005}
Arinstein, A.E. and Gitterman, M.: Random walks and anomalous diffusion in two-component random media. Physical Review E \textbf{72}, 021104 (2005).

% confined diffusion
\bibitem{Bickel_PhysicaA2007}
Bickel, T.: A note on confined diffusion. Physica A: Statistical Mechanics and its Applications \textbf{377}, pp. 24--32 (2005).

% confined diffusion
\bibitem{CalvoMunoz_PRE2011}
Calvo-Mu\~{n}oz, E.M., Selvan, M.E., Xiong, R., Ojha, M., Keffer, D.J., Nicholson, D.M. and Egami, T.: Applications of a general random-walk theory for confined diffusion. Physical Review E \textbf{83}, 011120 (2011).

%
%
% von Mises distribution
\bibitem{Codling_2008}
Codling, E., Plank, M., Benhamou, S.: Random walk models in biology. Journal of Royal Society Interface \textbf{5}, pp. 813–-834 (2008).

% von Mises distribution
\bibitem{Liu_JTB2015}
Liu, X., Xu, N.,  Jiang, A.: Tortuosity entropy: A measure of spatial complexity of behavioral changes in animal movement. Journal of Theoretical Biology \textbf{364}, pp. 197--205 (2015).

% von Mises distribution
\bibitem{Fofana_ProcB2017}
Fofana, A.M.,  Hurford, A.: Mechanistic movement models to understand epidemic spread. Philosophical Transactions of the Royal Society B: Biological Sciences \textbf{372}, 20160086 (2017).

% turning angle
\bibitem{Burov_PNAS2013}
Burov, S., Tabei, S.A., Huynh, T., Murrell, M.P., Philipson, L.H., Rice, S.A., Gardel, M.L., Scherer, N.F., Dinner, A.R.:  Distribution of directional change as a signature of complex dynamics. Proceedings of the National Academy of Sciences \textbf{110}, pp. 19689--19694 (2013).

% turning angle
\bibitem{Parisi_PRE2016}
Parisi, D.R., Negri, P.A., Bruno, L.: Experimental characterization of collision avoidance in pedestrian dynamics. Physical Review E \textbf{94}, 022318 (2016).

% turning angle
\bibitem{Liu_PRE2018}
Liu, P., Heinson, W.R., Sumlin, B.J., Shen, K.Y., Chakrabarty, R.K.: Establishing the kinetics of ballistic-to-diffusive transition using directional statistics. Physical Review E \textbf{97}, 042102 (2018).

% scaling behavior of ballistic motion
% \bibitem{Molinas-Mata_PRE1996}
% Molinas-Mata, P., Munoz, M.A., Martínez, D.O. and Barabási, A.L.: Ballistic random walker. Physical Review E \textbf{54}, pp. 968--971 (1996).

% scaling behavior of ballistic motion
\bibitem{Huang_PRE2002}
Huang, S.Y., Zou, X.W., Jin, Z.Z.: Directed random walks in continuous space. Physical Review E \textbf{65}, 052105 (2002).

% scaling behavior of ballistic motion
\bibitem{Visser_2006}
Visser, A.W., Kiørboe, T. Plankton motility patterns and encounter rates. Oecologia \textbf{148}, pp. 538--546 (2006).

% scaling behavior of ballistic motion
\bibitem{Smieszek_2009}
Smieszek, T., 2009. A mechanistic model of infection: why duration and intensity of contacts should be included in models of disease spread. Theoretical Biology and Medical Modelling \textbf{6}, 25 (2009).

% T_n, feature extraction 
\bibitem{Kowalek_PRE2019}
Kowalek, P., Loch-Olszewska, H., Szwabiński, J.: Classification of diffusion modes in single-particle tracking data: Feature-based versus deep-learning approach. Physical Review E \textbf{100}, 032410 (2019).

% T_n, feature extraction
\bibitem{Wagner_PLOS2017}
Wagner, T., Kroll, A., Haramagatti, C.R., Lipinski, H.G., Wiemann, M.: Classification and segmentation of nanoparticle diffusion trajectories in cellular micro environments. PLOS One \textbf{12}, e0170165 (2017).

% feature extraction
\bibitem{Pinholt_PNAS2021}
Pinholt, H.D., Bohr, S.S.R., Iversen, J.F., Boomsma, W., Hatzakis, N.S.: Single-particle diffusional fingerprinting: A machine-learning framework for quantitative analysis of heterogeneous diffusion. Proceedings of the National Academy of Sciences \textbf{118}, e2104624118 (2021).

\end{thebibliography}

\end{document}